\documentclass[]{article}

\usepackage{authblk}
\usepackage{blindtext}
\usepackage[skip=10pt plus1pt]{parskip}
\usepackage{amsmath, amsthm}
\usepackage{amssymb}
\usepackage{geometry} 
\usepackage{upgreek}
\usepackage{mathrsfs}
\usepackage{float}
\usepackage{subcaption}
\usepackage[toc,page]{appendix}
\usepackage{graphicx}
\usepackage{color, pifont}
\usepackage{graphicx} 
\usepackage{tikz}
\usepackage{enumitem}
\usepackage{tabularx}
\usepackage{multirow}
\newcolumntype{Y}{>{\centering\arraybackslash}X}
\usepackage{hyperref}       
\hypersetup{
  colorlinks   = true, 
  urlcolor     = purple, 
  linkcolor    = teal, 
  citecolor   = red 
}
\usepackage{url}

\newcommand*\colourcheck[1]{%
  \expandafter\newcommand\csname #1check\endcsname{\textcolor{#1}{\ding{52}}}%
}
\colourcheck{green}
\colourcheck{blue}

\title{A Mechanical Model for the Failure of Reconstructive Breast Implant Surgery Due to Capsular Contracture}
\author[1,*]{Yuqi Xiao}
\author[2]{Leah Edelstein-Keshet}
\author[3]{Alain Goriely}
\author[4]{Kathryn V Issac}
\affil[1, 2]{Department of Mathematics, University of British Columbia, Vancouver, BC, V6T 1Z2, Canada}
\affil[3]{Mathematical Institute, University of Oxford,
Woodstock Road, Oxford, OX2 6GG, United Kingdom}
\affil[4]{Department of Surgery, Faculty of Medicine, University of British Columbia, Vancouver, BC, Canada}
\affil[*]{ukixiao@math.ubc.ca}
\date{}
\begin{document}
\maketitle
\section*{Abstract}
Capsular contracture  is a pathological response to implant-based reconstructive breast surgery, where the ``capsule'' (tissue surrounding an implant) painfully thickens, contracts and deforms. It is known to affect breast-cancer survivors at higher rates than healthy women opting for cosmetic cosmetic breast augmentation with implants. We model the early stages of capsular contracture based on stress-dependent recruitment of contractile and mechanosensitive cells to the implant site. We derive a one-dimensional continuum spatial model for the spatio-temporal evolution of cells and collagen densities away from the implant surface.  Various mechanistic assumptions are investigated for linear versus saturating mechanical cell responses and cell traction forces. Our results point to specific risk factors for capsular contracture, and indicate how physiological parameters, as well as initial states (such as inflammation after surgery) contribute to patient susceptibility.

\textbf{Keywords:} Reconstructive Breast Surgery, Capsular Contracture, Mathematical Model, Mechanosensitivity, Physiological Feedback, Bistability

\section{Introduction}
\label{sec: intro}

Breast cancer is a common life-threatening disease that necessitates surgery with the possible addition of radiation and chemotherapy. Following surgical removal of cancerous tissue with a mastectomy, breast reconstruction is offered to restore a sense of normalcy and improve quality of life in survivorship. In implant-based breast reconstruction, a prosthetic implant is used to restore the breast mound. In normal healing, new scar tissue defined as the ``capsule'' forms around the implant surface, consisting of cells and extracellular matrix components A pathological result, \textit{capsular contracture} (CC) may develop in 15-40\% of cases with a higher rate of occurrence for breast cancer patients undergoing radiotherapy \cite{headon2015capsular}. The pathological capsule contracts and exerts pressure on the implant, within the span of months to years, resulting in chronic pain and deformity. CC usually mandates further surgery to remove the implant, a highly undesirable outcome. Our aim is to understand the causes and risk factors leading to this pathology.

Normal healing involves a cascade of cell recruitment, chemical signaling, cell contraction, remodeling of the extracellular matrix, and eventual resolution of the inflammatory response. There are reasons to believe that in CC, pathological tissue capsules develop as a result of abnormal healing with run-away or abnormal immune responses or abnormal fibrosis \cite{safran2021current}. Our purpose here is to explore the cellular-level root causes, with particular emphasis on cell mechanosensitivity.  

The precise mechanism underlying the development of capsular contracture remains elusive, but a few key observations have been made. First, the extent of the initial immune response varies from patient to patient, meaning that different surgical practices and clinical environment may lead to different outcomes. Distinct outcomes can also occur in the same patient (e.g., left vs right breast). Second, the time span of CC development ranges from months to years after a breast implant surgery, with many patients only developing symptoms years later. It has been suggested that this delayed response could result from sudden immune insults, such as bacterial infection \cite{hanna2023cutibacterium}; micro-tears of the implant due rupture \cite{dancey2012capsular}; or even a COVID-19 vaccination \cite{restifo2021case}, leading to the activation of immune cells in the proximity to the implant.

It is challenging to study CC etiology, as histological and mechanical data are typically only collected when pathological capsules are extracted in follow-up surgery. Clinical experiments are costly and can be difficult to conduct, as monitoring the early stages requires invasive measurement techniques. Therefore, mathematical modeling is a cost efficient way to test hypotheses and advance our fundamental understanding of the mechanisms responsible for CC.

In a preliminary mathematical model of CC from this group \cite{dyck2024models}, we described and simplified the complex cellular interactions in CC to interactions of several cell types that produce the extracellular matrix (ECM) collagen. Collagen density was used as a surrogate for tissue ``stiffness'', assumed to create feedback amplifying the inflammatory cycle. The model neglected the deformation, stress, and cellular mechanosensitivity, factors that have been recognized as important determinants of cellular dynamics. Our aim in this paper is to derive and investigate a mechanochemical model that takes these factors into account. 

Mechanical models of living tissue were proposed in \cite{murray2003Book}, and used in \cite{tranquillo1993mechanistic} to study wound healing. Villa et al.~\cite{villa2021mechanical} use similar ideas to investigate patterns of cell and stress distribution that arise from distinct constitutive relations for tissue mechanical properties. Such mechanical considerations were neglected in our earlier model of CC \cite{dyck2024models}. Here, we build upon that model to include interactions and spatial distributions of cells, collagen, and stress. We derive a minimal partial differential equation (PDE) model, estimate its parameters from the literature, and examine its behavior. 

We address the following questions: 
(1) What are the mechanochemical factors and interaction at the cellular level that drive the development of CC?
(2) What patient factors influence the risks and severity of CC outcomes? 
(3) What are potential treatment strategies targeting these risk factors? 

\section{Model derivation}
\label{sec: derivation}

It is known that the insertion of an implant triggers a foreign body response (FBR) (See \cite{dyck2024models, safran2021current} for a review.) At early stages of FBR, macrophages and fibroblasts are recruited to the site \cite{pakshir2018big}. Fibroblasts and myofibroblasts produce collagen \cite{li2011fibroblasts, klingberg2013myofibroblast}, one of the most abundant fibrous protein in human breast tissue \cite{frantz2010extracellular}. High collagen density found in pathological capsules is often linked to more severe cases of contracture \cite{mcdougall2006fibroblast}. Collagen is part of the ECM, providing structural scaffold for for cell movement and cell forces. Fibroblasts and myofibroblasts are contractile cells that can pull on and create tension along ECM fibers. 
It has been found that stress due to contracting fibroblasts can recruit remote macrophages \cite{pakshir2019dynamic}. 

\begin{figure}[ht]
    \centering
    \includegraphics[width=0.6\linewidth]{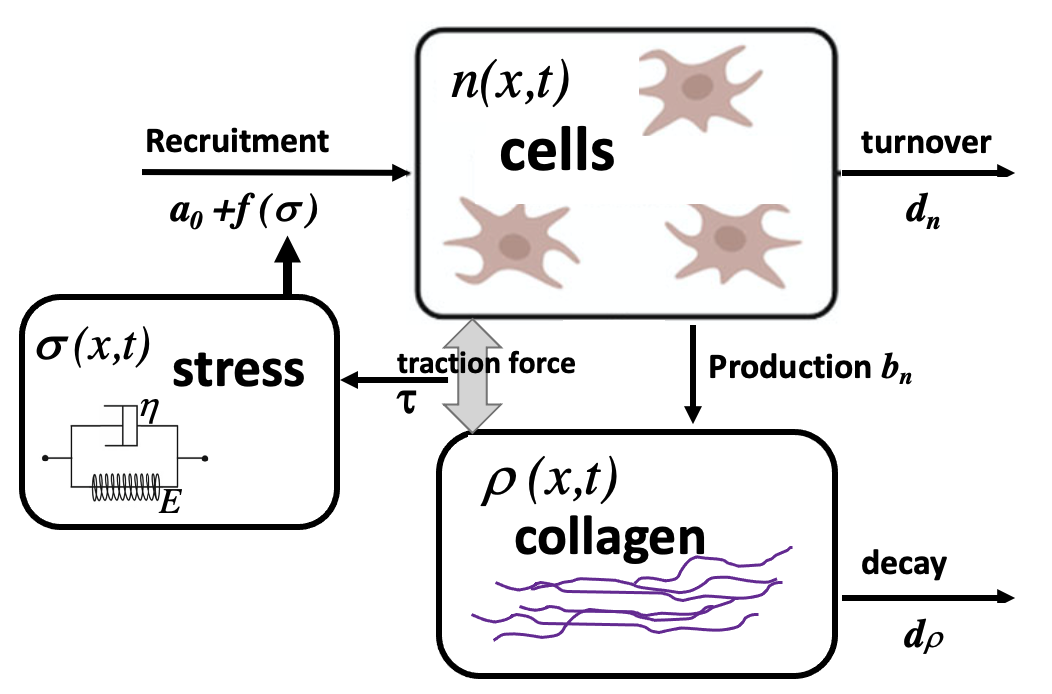}
    \caption{Schematic diagram for the model, adapted from \cite{dyck2024models}. Cells (density $n(x,t)$) produce collagen fibers (density $\rho(x,t)$) and create stress ($\sigma(x,t)$) by exerting traction forces ($\tau(n,\rho)$) on those fibers. Stress (based on Kelvin-Voight constitutive relation) amplifies further recruitment of cells, resulting in positive feedback that we hypothesize could act as a  driver of capsular contracture.
    }
    \label{fig:model-schematic}
\end{figure}

To highlight the role of mechanical feedback, we  consider here a simple
model with a single force-producing cell type, collagen production, and mechanosensitive cell recruitment. This feedback mechanism is depicted in Figure \ref{fig:model-schematic}. In an effort to investigate the role of this feedback mechanism in CC formation, we model different assumptions about (1) stress-dependent cell recruitment; (2) cell traction forces. 

\begin{figure}[H]
    \centering        
    \includegraphics[width=0.7\textwidth]{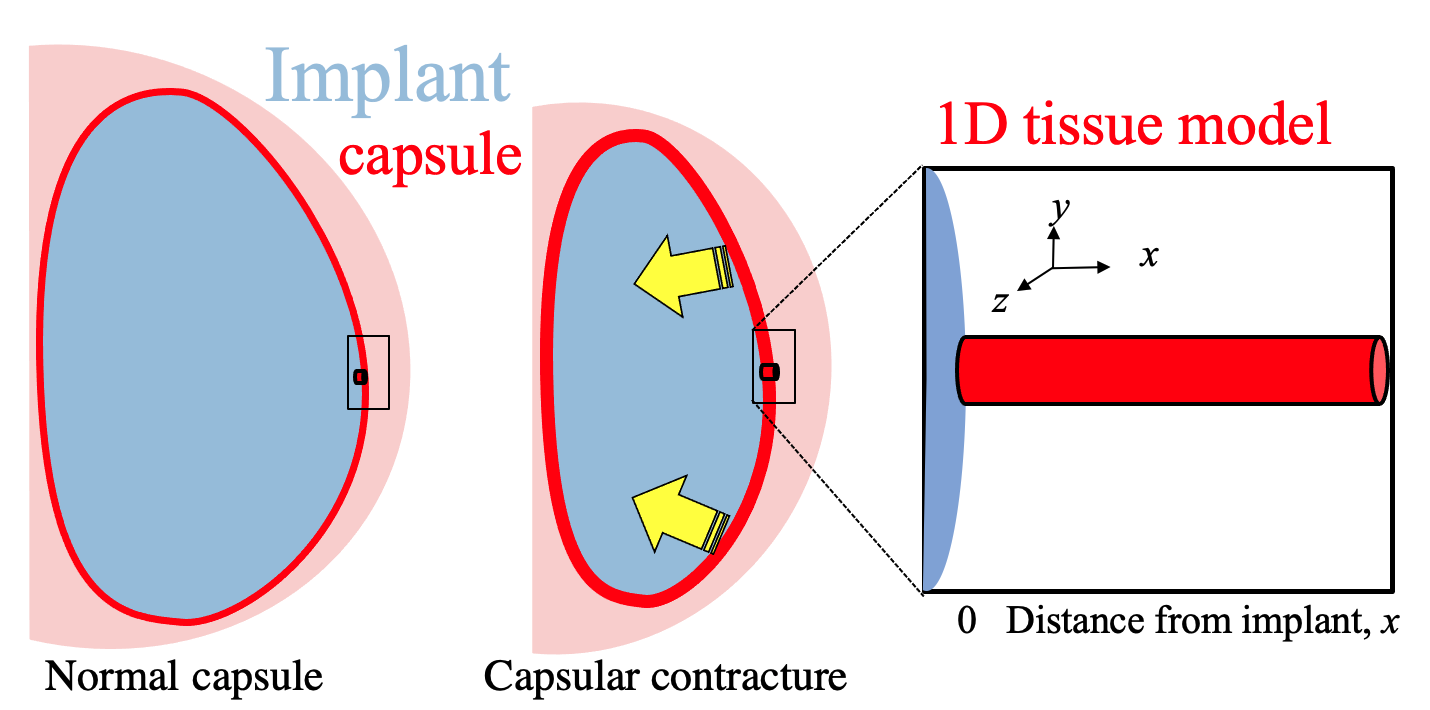}
    \caption{Left: schematic diagrams of normal capsule and CC, depicting thickening of the capsule and deformation in the latter. Right: A simplified 1D model 
    geometry. We consider a thin 1D tube of capsule tissue along the $x$-axis, where $x$ is the distance of tissue from the implant site, $0\leq x\leq L$.}
    \label{fig:geometry}
\end{figure}

Many biomechanical models of tissue are posed in 2D or 3D geometry, in order to account for full shape deformation. But here, our main interest resides in the cell-level mechanisms, and we do so in the simplified 1D geometry shown in Figure~\ref{fig:geometry}. In our
1D domain, the spatial dimension represents the distance away from the implant. While it is clear that the stress field is a three-dimensional field that needs to be specified in the full geometry, our interest here is to identify the effect of longitudinal contraction away from the capsule, which justifies our choice. 

In the following derivations, we let $n(x,t), \rho(x,t)$ be densities of cells and collagen, and we denote by $u(x,t), v(x,t), \sigma(x,t)$ 
the displacement of tissue from its original unstrained location, the rate of change of displacement (i.e. velocity), and the axial stress, at distance $x$ from the implant site at time $t$, respectively.

\subsection{Dynamics of cells}

We assume that cells have a basal recruitment rate $a_n$, a turnover rate proportional to cell density $d_n n$, and a random motility coefficient $D_n$. Importantly, we examine the hypothesis that cells are recruited by stress at some rate, $f(\sigma)$, where 
\begin{equation}
    f(\sigma) = a_0\sigma + a_1\frac{\sigma^{k_1}}{\sigma_0^{k_1}+\sigma^{k_1}}.
\end{equation}

We will consider the cases where (i) there is no stress-related recruitment of cells ($a_0 = a_1 = 0$), (ii) cells are recruited by stress at a rate linearly proportional to stress ($a_0 \not= 0, a_1 = 0$), and (iii) cells are recruited by stress at a nonlinear, saturating rate (a Hill function; $a_0 = 0, a_1 \not=0$, with $\sigma_0\ge 0$ a constant depicting the level of stress associated with 50\% of the maximal stress-induced recruitment rate; the Hill coefficient $k_1>0$ governs the steepness of the response - for large $k_1$ the stress response switches abruptly to its full intensity as soon as stress is above the threshold level $\sigma_0$). The terms $a_0, a_1$ here could also represent stress-dependent cell activation, turning quiescent cells into actively contractile cells \cite{lewalle2024cardiac}.

Mass conservation for cells on the 1D domain results in the PDE,
\begin{equation}
    \frac{\partial n}{\partial t} = \frac{\partial}{\partial x}(D_n\frac{\partial n}{\partial x} - nv) + a_n + f(\sigma) - d_nn,
    \label{eq:cells}
\end{equation}
where $v = u_t$ is the rate of deformation. This term is analogous to a transport velocity, as it results in a net rate of displacement of the tissue, which carries cells and collagen with it.

\subsection{Dynamics of collagen}
We assume a collagen production rate by cells $b_nn$, a collagen decay rate $d_{\rho}\rho$, a random reorganization rate $D_{\rho}$, and a transport flux due to the distortion of tissues, similar to cells. Putting these elements together, we arrive at the equation for the evolution of collagen,
\begin{equation}
    \frac{\partial \rho}{\partial t} = \frac{\partial}{\partial x}(D_{\rho}\frac{\partial \rho}{\partial x} - \rho v) + b_nn - d_{\rho}\rho.
    \label{eq:collagen}
\end{equation}

\subsection{Force equation for the cell-collagen system}
Following Murray et al. \cite{murray2003mathematical}, we consider the breast tissue as a linear viscoelastic material with low Reynolds number (negligible inertial force), and we assume the tissue to be in mechanical equilibrium. In contrast to Murray et al. \cite{murray2003Book}, where a layer of cells adheres to a substrate, our 1D tissue has no significant adhesion forces that act as external body forces. Adhesion to the implant is represented by our boundary conditions (BC's) at $x=0$. Then the total stress on the tissue $\sigma(x,t)$, satisfies the 1D force-balance equation 
\begin{equation}
    \frac{\partial \sigma}{\partial x} = 0.
    \label{eq:force-balance}
\end{equation}
This assumption implies that stress is spatially uniform in the 1D tissue domain. This stress can be decomposed into $\sigma_{cells}$, the stress contributed by cells, and $\sigma_{ECM}$, the stress contributed by the collagenous ECM:
\begin{equation}
    \sigma = \sigma_{cells} + \sigma_{ECM}.
    \label{eq:total-stress}
\end{equation}
We assume that cells generate stress by pulling on collagen fibers, in the form 
\[
    \sigma_{cells} = g(n,\rho) = \tau_1 n\rho + \tau_2\frac{n^{k_2}}{n_0^{k_2} + n^{k_2}}\rho.
\]
Here the  $g(n,\rho)$ is the traction force as a function of cell and collagen density.
We will consider two cases where: (i) the traction force is linearly proportional to both cell and collagen density ($\tau_1 \not=0, \tau_2=0$); (ii) the traction force is linearly proportional to collagen density, and is a nonlinear saturating function of cell density,  taken to be a Hill function ($\tau_1 = 0$, $\tau_2 \not=0$). The case of $k_2>1$ represents cell cooperativity, that is, the switch to greater pulling activity when cell density exceeds some critical level $n_0>0$. Such effects is associated with local cell signaling and mechanosensitivity.

To close the system, we need to make some constitutive assumption about how tissue deformation results from mechanical forces. Living tissue can be modelled as a viscoelastic material. A constitutive relation is used to describe the stress-strain relationships of viscoelastic material such as the ECM. In \cite{villa2021mechanical}, six variants of the constitutive relations were investigated. Here we adopt the Kelvin-Voigt case, a common and well-supported assumption for tissue \cite{catheline2004measurement,goriely17}. The corresponding stress-strain constitutive equation is 
\[
    \sigma_{ECM} = E\epsilon + \eta \epsilon_t,
\]
where $E$ and $\eta$ denotes the Young's modulus and the viscosity coefficient of the collagenous ECM, respectively, and $\epsilon$ is the strain, given simply in  1D by
\[
\epsilon = u_x,
\] so that, we have 
\[
    \sigma_{ECM} = E u_x + \eta u_{tx}.
\]
With these assumptions, we can combine equations \eqref{eq:force-balance} and \eqref{eq:total-stress}, to obtain 
\[
 Eu_{xx} + \eta u_{txx} + g(n,\rho)_x = 0.
\]
This single PDE links derivatives of the displacement to the traction forces created by cells contracting and pulling on collagen.

\subsection{The mechano-CC model}
Combining all equations, we arrive at the \textit{mechano-CC} model. 
\begin{subequations}
    \begin{align}
    \text{Cells}\qquad      &\frac{\partial n}{\partial t} = D_n\frac{\partial^2 n}{\partial x^2} - \frac{\partial}{\partial x}(nv) + a_n + 
    a_0\sigma + a_1\frac{\sigma^{k_1}}{\sigma_0^{k_1}+\sigma^{k_1}} - d_nn, \\
    \text{Collagen}\qquad    &\frac{\partial \rho}{\partial t} = D_{\rho}\frac{\partial^2 \rho}{\partial x^2} - \frac{\partial}{\partial x}(\rho v) + b_nn - d_{\rho}\rho,\\
    \text{Stress}\qquad    &\sigma = E\frac{\partial u}{\partial x} + \eta \frac{\partial^2 u}{\partial t \partial x} + \tau_1 n\rho + \tau_2\frac{n^{k_2}}{n_0^{k_2} + n^{k_2}}\rho, \\
    \text{Force balance}\qquad      &\frac{\partial \sigma}{\partial x} = 0.
    \end{align}
    \label{eqn:full_model}
\end{subequations}
The parameters in the mechano-CC model are all non-negative. See definitions and units in Table~\ref{tab:paramsMeaning}.

\subsection{Unstressed homogeneous steady state}
Consider the case of unstressed tissue where $\sigma=0$ everywhere. Then \eqref{eqn:full_model} has a homogeneous steady state (HSS) where cells and collagen are at the levels
\begin{equation}
\label{eq:HSS}
    n_{SS}= a_n/d_n, \quad \rho_{SS}= n_{SS} (b_n /d_\rho) =a_n b_n/ d_n d_\rho. 
\end{equation}
We anticipate that far away from the implant, the cell and collagen density would be given by these unstressed steady state levels, leading to our boundary conditions.

\subsection{Boundary conditions}

We use the following boundary conditions for the mechano-CCl model. We assume that the implant is impermeable to cells and collagen, implying no flux of cells and collagen at $x = 0$. We also assumed that cell and collagen densities drop to their normal unstressed HSS level far away from the implant ($x = L$). We assume that the displacement of tissue is zero at both boundaries. 
\begin{subequations}
{
\[
    \begin{array}{ll}
    \text{Cells}\qquad &\frac{\partial n}{\partial x}(0,t) = 0, \quad n(L,t) = n_{SS}, \\
    \hfill \\
    \text{Collagen}\qquad &\frac{\partial \rho}{\partial x}(0,t) = 0, \quad \rho(L,t) = \rho_{SS}, \\
    \hfill \\
    \text{Displacement}\qquad &u(0,t) = u(L,t) = 0.
    \end{array}
\]
} 
\label{eqn:bcs}
\end{subequations}
Where $n_{SS}, \rho_{SS}$ are defined in \eqref{eq:HSS}. These are also validated and compared against literature values in section \ref{sec: params}. We will also explore the possible effect of other BCs for $u$ later.

\section{Scaling and approximating the model}
\label{sec: analysis}

\subsection{Dimensional analysis}

We scale space, time, cell density, collagen density, and stress,
\[
    x^* = \frac{x}{L},\quad t^* = \frac{t}{T},\quad n^* = \frac{n}{N},\quad \rho^* = \frac{\rho}{\varrho},\quad u^* = \frac{u}{L},\quad \sigma^* = \frac{\sigma}{S}.
\]
We choose to scale time by the cell residence time $T=1/d_n$, space by the distance of random cell motion over this characteristic time, $L = \sqrt{D_nT}$. Cell density is scaled by its unstressed homogeneous steady state,$N=n_{SS}$.
Collagen density is scaled by its unstressed HSS multiplied by the ratio of turnover rate of collagen to cells $\varrho = \rho_{SS}\cdot(d_{\rho} /d_n).$
The resulting choices for scales is then
\[
    T = \frac{1}{d_n},\quad L =  \sqrt{\frac{D_n}{d_n}},\quad N = \frac{a_n}{d_n},\quad \varrho  = \frac{b_na_n}{d_n^2}.
\]
Note that we did not choose to scale $x$ by the domain length since this can vary greatly between patients. Additionally, cho

Transforming the variables and simplifying the PDEs leads (after dropping the *'s) to the dimensionless model
\begin{subequations}
  \label{eq: non-dimensioned_general_model}
    \begin{align}
        &n_t = n_{xx} - (nv)_x + 1 + f(\sigma) - n, \\
        &\rho_t = D_2{\rho}_{xx} - (\rho v)_x + n - d_2\rho,\\
        &\sigma = E_1u_x + \eta_1u_{tx} + g(n,\rho), \\
        &\sigma_x = 0.
    \end{align}
where we later consider several variants, namely
\begin{equation}
        f(\sigma) = \begin{cases}
        f_1=0, \quad \text{(None)}, \quad  \text{ for } a_0 = a_1 = 0,\\
        f_2=\sigma, \qquad \text{(Linear)}, \text{ for } a_0 \not = 0, a_1 = 0,\\
        f_3=a\frac{\sigma^{k_1}}{1+\sigma^{k_1}},  \quad \text{(Hill)}, \text{ for } a_0 = 0, a_1 \not= 0.
    \end{cases}
\end{equation}

and 
\begin{equation}
    g(n,\rho) = \begin{cases}
        g_1=\tau'_1 n\rho , \quad \text{(Proportional)} \text{ for $\tau'_1 \not= 0, \tau'_2 = 0$ } ,\\
        g_2= \tau'_2\frac{n^{k_2}}{N_0^{k_2} + n^{k_2}}\rho,  \quad \text{(Hill)}, \quad \text{ for $\tau'_1 = 0, \tau'_2 \not= 0$} ,\
    \end{cases}
\end{equation}
\end{subequations}

The new model parameters, all dimensionless, are now:
\begin{gather*}
    a = \frac{a_1}{a_n}, \qquad D_2 = \frac{D_{\rho}}{D_n}, \qquad d_2 = \frac{d_{\rho}}{d_n}, \qquad E_1 = \frac{E}{S}, \qquad \eta_1 = \frac{\eta d_n}{S},\\
    \qquad \tau'_1 =\frac{\tau_1\varrho N}{S} = \frac{\tau_1 a_n^2b_n}{S d_n^3}, \qquad \tau'_2 =\frac{\tau_2\varrho}{S}= \frac{\tau_2a_nb_n}{S d_n^2}, \qquad N_0 = \frac{n_0}{N}=\frac{n_0d_n}{a_n},
\end{gather*}
where 
\[
    S = \begin{cases}
        E \qquad\text{ for } a_0 = a_1 = 0,\\
        \frac{a_n}{a_0} \qquad\text{for } a_0 \not = 0, a_1 = 0,\\
        \sigma_0 \qquad\text{ for } a_0 = 0, a_1 \not= 0.
    \end{cases}
\]
The dimensionless parameters have the following interpretations: $a$ is a fold-multiple for the stress-induced amplification of basal cell recruitment. $D_2$ is a ratio of random rearrangement/dispersal of collagen to cells. $N_0$ is a scaled threshold cell density at which cooperative cell traction forces are switched on (expressed as a fold-multiple of the unstressed HSS cell density). $d_2$ is the  ratio of collagen to cell turnover rates. The mechanical parameters are all scaled by a reference stress level that is model dependent. Hence, $E_1, \eta_1$ are scaled elastic modulus and viscosity. $\tau_1'$ is  the ratio of two stresses: traction force produced at the low (unstressed HSS) cell and collagen densities relative to the reference stress. Similarly, $\tau_2'$ is a ratio of maximal traction force (at high cell densities and HSS collagen) to the reference stress level. 

\subsection{Approximating an analytic solution}
In this section, we consider the following version of our model \eqref{eq: non-dimensioned_general_model} by setting $a_0 = 0$, $\tau'_1 = 0$.
The  model \eqref{eq: non-dimensioned_general_model} is nonlinear and not analytically tractable as is, but a few approximations can be made to reveal its basic behavior. First, we look for steady state solutions, satisfying $n_t = \rho_t = u_t = 0$, leading to
\begin{subequations}
    \begin{align}
        &n_{xx} + 1 + a\frac{\sigma^{k_1}}{1+\sigma^{k_1}} - n = 0 , \\
        &D_2{\rho}_{xx} + n - d_2\rho = 0,\\
        &\sigma - E_1u_x - \tau'_2\frac{n^{k_2}}{N_0^{k_2} + n^{k_2}}\rho = 0, \\
        &\sigma_x = 0.
    \end{align}
    \label{eqn:BVP_equations}
\end{subequations}
We consider the case of a sharp response, namely when the Hill coefficient, $k_1$, is large enough so that the Hill function can be approximated by a switch-like (Heaviside) function
\begin{equation}
\label{eq:sharp}
     \frac{\sigma^{k_1}}{1+\sigma^{k_1}} =
     \text{Heaviside}(\sigma-1)=
     \begin{cases}
        0 \qquad \text{for } \sigma < 1, \\
        1 \qquad \text{for } \sigma \ge 1
    \end{cases}
\end{equation}
In this case, the model can be solved explicitly, resulting in the cell density distribution
\begin{equation}
\label{eq:approxn(x)}
n(x)=\left\{
\begin{array}{ll}
     1 + \frac{n_{SS}-1}{e+e^{-1}}(e^x+e^{-x}),\quad &\text{normal},   \\
    &\\
       1 + a + \frac{n_{SS}-1-a}{e+e^{-1}}(e^x+e^{-x}),\quad &\text{elevated}.
\end{array} \right.
\end{equation}
We can plot these solutions and compare them against the numerical results produced by a MATLAB BVP solver (detailed in Appendix \ref{appendix: BVP}).

\begin{figure}[H]
    \centering
    \includegraphics[width=0.7\linewidth]{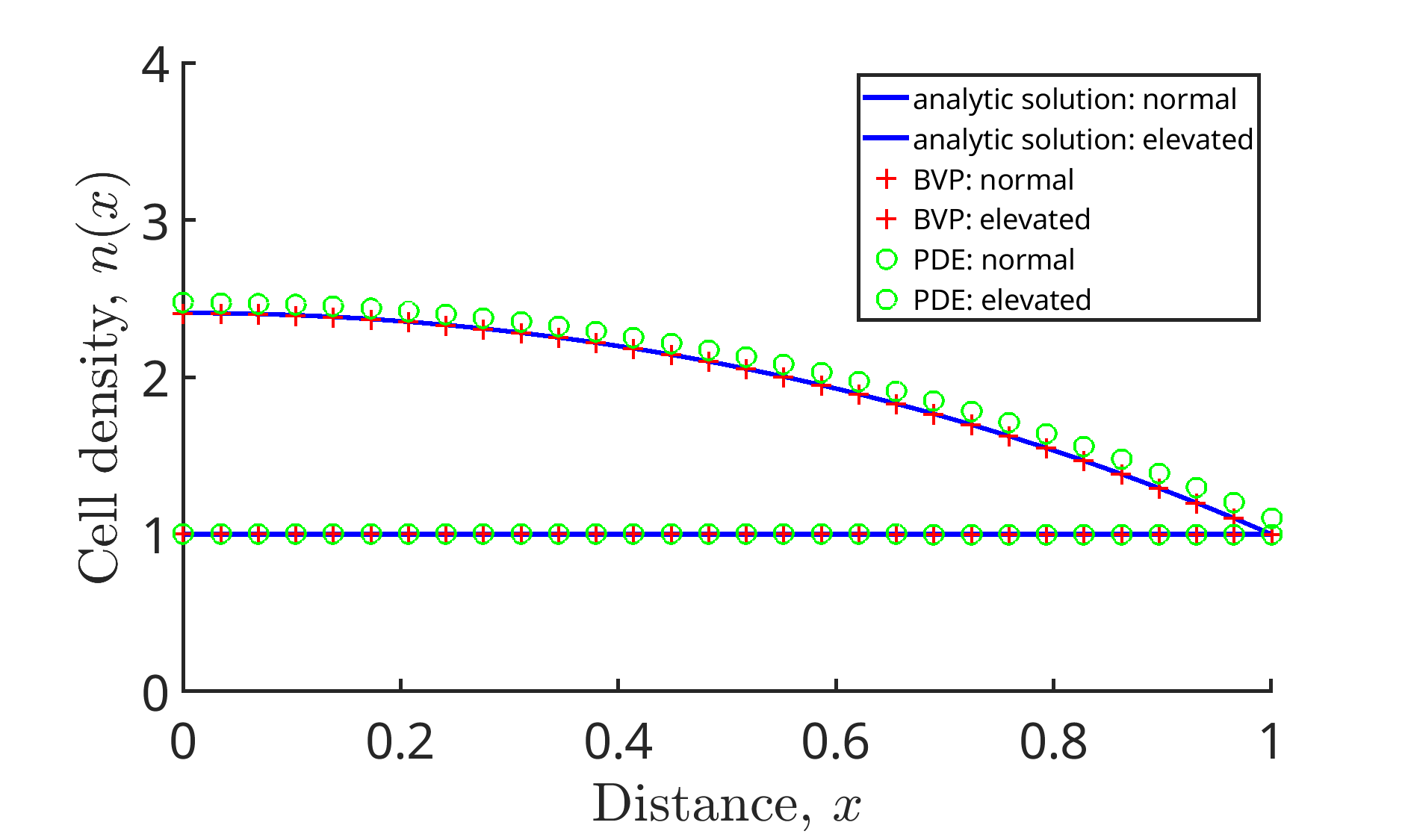}
    \caption{\textbf{Numerics validation: }Typical steady state cell density (vertical axis) versus distance from implant $x$ satisfying \eqref{eqn:BVP_equations} with the 
``sharp switch'' stress recruitment approximation \eqref{eq:sharp}. Solid blue: analytic solution. Red crosses: MATLAB BVP solutions of the model. Green circles: Steady state solutions predicted by our full time-dependent PDE simulations.
For the given set of parameters (see Appendix and \eqref{eqn:ICs-params-BVP}), a uniform low steady state coexists with a steady state of elevated cell density.
MATLAB simulation results generated with the initial conditions, initial guesses and parameters in Equations \eqref{eqn:ICs-params-BVP}.}
    \label{fig:analytic_bvp_compare}
\end{figure}

\section{Parameter estimation}
\label{sec: params}

In many cases, the estimation of parameters is restricted by the availability of data in existing literature. We collected quantitative information for various cell types known to be implicated in CC, including macrophages, fibroblasts and myofibroblasts to create a composite cell parameter regime. Many species, sources and types of studies were used to gain some preliminary estimates. From such studies, we assembled the values for key parameters.

Some preliminary parameters in the full model 
\eqref{eqn:bcs} are important for our dimensionless model \eqref{eq: non-dimensioned_general_model}. Because we are interested in cell-scale events, we converted all quantities to units of $\mu$m, and seconds. We summarize the values for these parameters in Table \ref{tab:paramsValues} and explain briefly how they are estimated. Fuller details are available in Appendix~\ref{appendix: ParamDetails}.

\begin{table}[H]
    \centering
    \renewcommand{\arraystretch}{1.4}
    \begin{tabularx}{\textwidth}{|c|l|X|l|} \hline
        Parameter &Meaning & Estimated value & Source \\ \hline 
        $n_{SS}$ 
        & Typical HSS cell density
        & $10^{-7}$ cells/$\mu$m$^3$ & \cite{wigginton2001model} \\ \hline

        $\rho_{SS}$ 
        &Typical HSS collagen density
        & $1.3$-$3\times 10^{-12}$mg$/\mu$m$^3$ & \cite{provenzano2008collagen} \\ \hline
        
        $a_n$ 
        & Basal cell recruitment rate
        & 0.76-5.2$\times 10^{-14}$cells/($\mu$m$^3\cdot$ s) 
        & \cite{wigginton2001model, katz1991human} \\ \hline

        $d_n$ 
        & Cell turnover rate
        &   0.35-10$\times 10^{-6}$/s & \cite{agaiby1999immuno, bartha2022mathematical, baum2011fibroblasts, frangogiannis2000myofibroblasts} \\ \hline

        $b_n$ 
        & Collagen secretion rate
        & 0.5-1.4$\times10^{-11}$mg/cells/s & \cite{illsley2000increased} \\ \hline
        
        $d_{\rho}$ 
        &Collagen decay rate
        & $1.6 \times 10^{-6}$/s & \cite{cleutjens1995regulation} \\ \hline

        $E$ 
        & Tissue elastic modulus
        & 2.34-4.16 $\times 10^3$ Pa & \cite{samani2007elastic} \\ \hline

        $\eta$ 
        & Tissue viscosity coefficient
        & 16.4-29.1 Pa$\cdot$s & \cite{samani2007elastic, liu2023real} \\ \hline

        $\tau_1$ 
        & Cell traction parameter
        & 1-3$\times 10^{22} \,
        \text{Pa}\cdot {(\text{cells}/\mu\text{m}^3)^{-1}\cdot(\text{mg}/\mu\text{m}^3)}^{-1}$ 
        & \cite{chen2007alpha} \\ \hline

        $\tau_2$ 
        & Cell traction parameter
        & 1-3$\times 10^{15} \,
        \text{Pa}\cdot{(\text{mg}/\mu\text{m}^3)^{-1}}$
        & \cite{chen2007alpha} \\ \hline
    \end{tabularx}
    \caption{Estimated values of parameters for the model \eqref{eqn:full_model} obtained from the literature.}
    \label{tab:paramsValues}
\end{table}

\textbf{Cell densities and dynamics:} 
Our model \eqref{eqn:full_model} contains a single cell type that depicts a composite of the properties of various cells implicated in CC. Actual cells that participate in the process of healing include macrophages, fibroblasts, and myofibroblasts. We used data for each, where available, to estimate ``cell parameters''.
Quantification of macrophage density, recruitment, and turnover was obtained from modeling studies of immunological responses in tuberculosis \cite{wigginton2001model}, and from non small cell lung cancer (NSCLC) \cite{bartha2022mathematical}. Some data was also available in \cite{agaiby1999immuno} for cell response to skin injury. Recruitment of myofibroblast in response to macrophages was obtained from \cite{katz1991human}. We found values for myofibroblast dynamics in cardiac injury in \cite{frangogiannis2000myofibroblasts}.

\textbf{Collagen density and dynamics:} We found sources for typical collagen density in murine mammary tissues in \cite{provenzano2008collagen}. Collagen production by dermal fibroblasts was given in 
\cite{illsley2000increased}, from which we 
could estimate a typical rate of collagen secretion by cells. The timescale of collagen turnover is based on recovery from cardiac injury data in
\cite{cleutjens1995regulation}.

\textbf{Tissue properties and forces:}
The elastic modulus for normal and cancerous breast tissue and tissue phantoms was available from \cite{liu2023real}. Cell traction forces were estimated based on corneal fibroblasts studied by \cite{chen2007alpha}.

Further details about the derivation of dimensionless parameters, and selection of appropriate initial conditions are explained in Appendix~\ref{sec:AppdxDim'lessPar}.


\section{Numerical simulation}
\label{sec: numerical_sim}

Following the approach of Villa et al \cite{villa2021mechanical}, we discretize the equations with respect to the spatial variable $x$ using finite differences, and solve the model as a system of ODEs with respect to time (method of lines). We use the implicit MATLAB ODE solver ODE15i to solve the resultant system of ODEs. The final outcomes are compared with approximate analytic solutions and numerical approximation of BVP results. To remove numerical instabilities, a small regularizing term is added to  the collagen equation. With this adjustment, detailed in  Appendix \ref{appendix: numerics}, the method is stable and accurate over a large range of parameters.

\subsection{Simulation results}

The dimensionless model \eqref{eq: non-dimensioned_general_model} has six variants depending on three choices for stress-related recruitment of cells and two choices for traction force. We summarize these variants in the Table~\ref{tab:Variants}. 
\begin{table}[H]
    \centering
    \renewcommand{\arraystretch}{1.2}
    \begin{tabularx}{0.85\textwidth}{|c|Y|Y|Y|}
    \hline
    Model Variant & Stress-related cell recruitment term & Traction force term & Model prediction and Figure number \\ \hline
        
    v-NL & 0 & Linear$(n\rho)$ & Healthy, Fig.~\ref{fig:v01-02} \\ \hline
        
    v-NH & 0 & Hill$(n)\rho$ & Healthy \\ \hline

    v-LL & Linear$(\sigma)$ & Linear$(n\rho)$ & CC, Figs.~\ref{fig:v11-time}, \ref{fig:v11} \\ \hline

    v-LH & Linear$(\sigma)$ & Hill$(n)\rho$ & CC, Fig.~\ref{fig:v12} \\ \hline

    v-HL & Hill$(\sigma)$ & Linear$(n\rho)$ & CC, Fig.~\ref{fig:v21} \\ \hline

    v-HH & Hill$(\sigma)$ & Hill$(n)\rho$ & Bistable, Figs.~\ref{fig:v22}, \ref{fig:v22-bistable}\\ \hline
    \end{tabularx}
    \caption{Model variants and the predicted outcome based on simulations of the model~\eqref{eq: non-dimensioned_general_model}. The notation ``Hill$(n)\rho$'' means a Hill function in cell density $n$, and proportional to collagen density $\rho$. The figures corresponding to the above variants are indicated.}
    \label{tab:Variants}
\end{table}

\begin{figure}[H]
    \centering
    \includegraphics[width=0.9\textwidth]{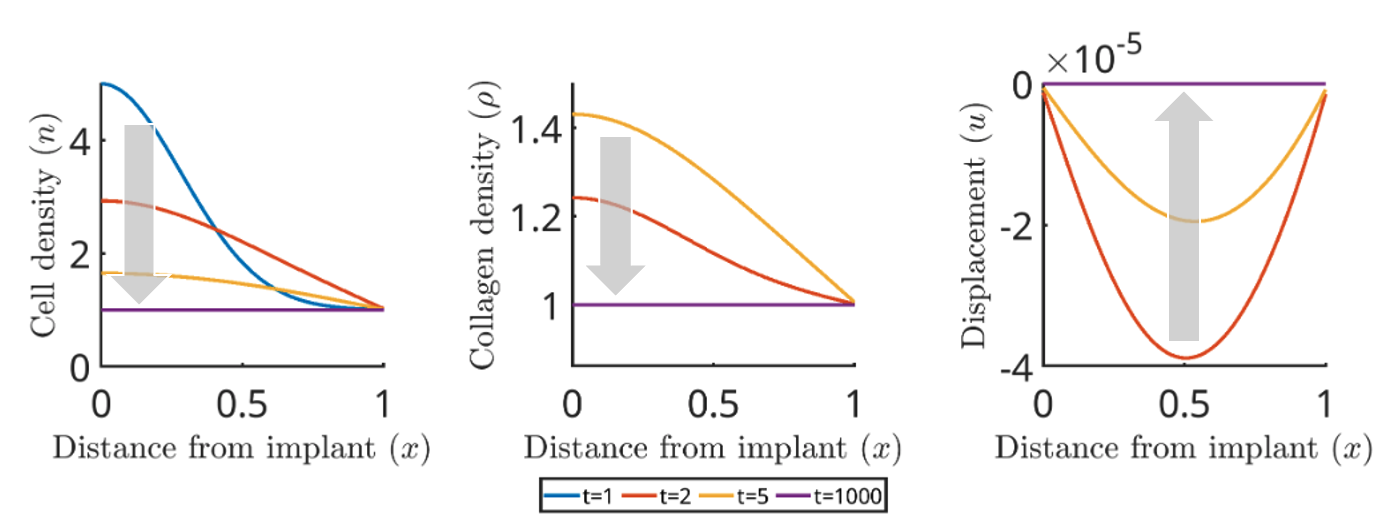}

    \caption{\textbf{Healthy resolution in absence of stress-dependent cell recruitment.} Time progression of cells and collagen densities, and displacement in model variant v-NL. Note that the collagen density and displacement at $t=1$ overlaps with the solution at $t=1000$, and is therefore not visible on the graph. All variables resolve to a spatially homogeneous steady state (HSS).  Parameters and initial conditions given in \eqref{eqn: ICs-params-v01-02}. 
    }
    \label{fig:v01-02}
\end{figure}

We run the two variants with no stress-related cell recruitment (v-NL and v-NH) with various initial conditions for cells, representing various initial immune response severity. Figure \ref{fig:v01-02}, shows the time progression of cells, collagen, and displacement vs distance from the implant. (Grey arrows depict time progression.)
Simulation results were consistent across a range of initial conditions over these two model variants, and over a range of reasonable values for $\tau'_1$ and $\tau'_2$. The final outcomes of these models consistently shows a spatially uniform cell and collagen density profile. We characterize these results as ``healthy outcomes'' in Table~\ref{tab:Variants} since neither cells nor collagen is elevated. These model variants do not account for progression to CC. While cells can contract and cause traction forces and stressed tissue, that stress does not amplify cell recruitment in v-NL and v-NH.

\subsubsection{Only pathological outcomes observed in the absence of multiple cooperative feedback mechanisms}

\begin{figure}[H]
    \centering
    \includegraphics[width=0.8\textwidth]{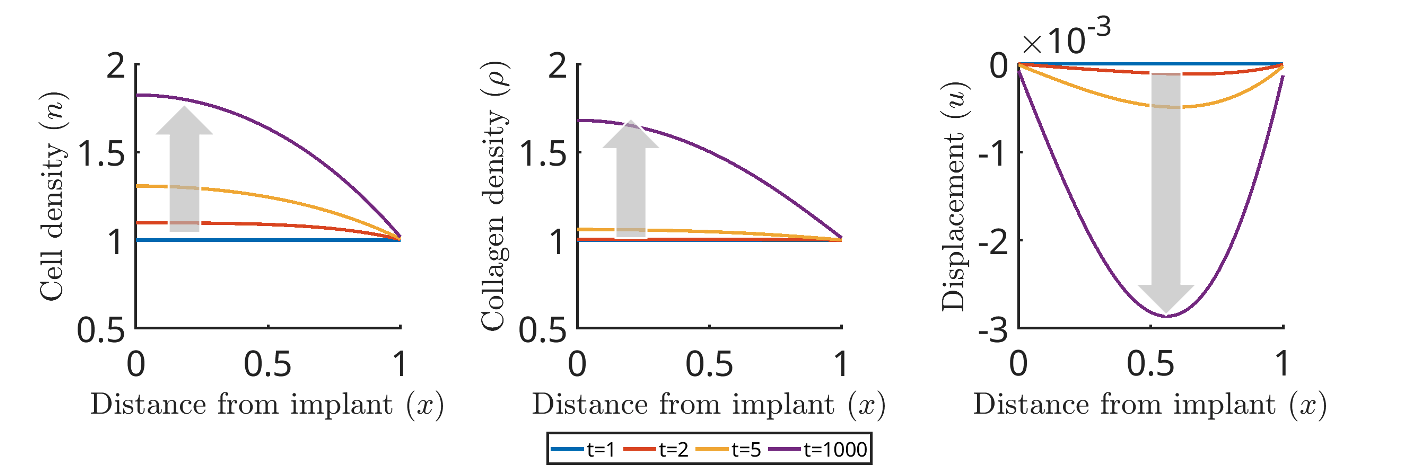}
    \caption{\textbf{Pathological progression in the absence of saturating feedback mechanisms.} Time progression of cells and collagen densities, and displacement in model v-LL. Parameters and initial conditions given in \eqref{eqn: ICs-params-v11}.
    }
    \label{fig:v11-time}
\end{figure}

Next, we  explore the model variants with stress-related cell recruitment proportional to stress, for both types of traction force assumptions (v-LL and v-LH). Figure \ref{fig:v11-time} displays the results for variant v-LL. In the range of parameters of interest, we find that the model always leads to a pathological outcome with elevated cells and collagen, regardless of initial conditions (close to the HSS). This suggests that a linear feedback relationship between cell-induced tissue stress and cell recruitment to the tissue is too extreme, leading to a bias towards the CC outcome. The same results are observed in the time behavior of variants v-LH and v-HL (not shown). 

\begin{figure}[H]
    \centering
    \includegraphics[width=\textwidth]{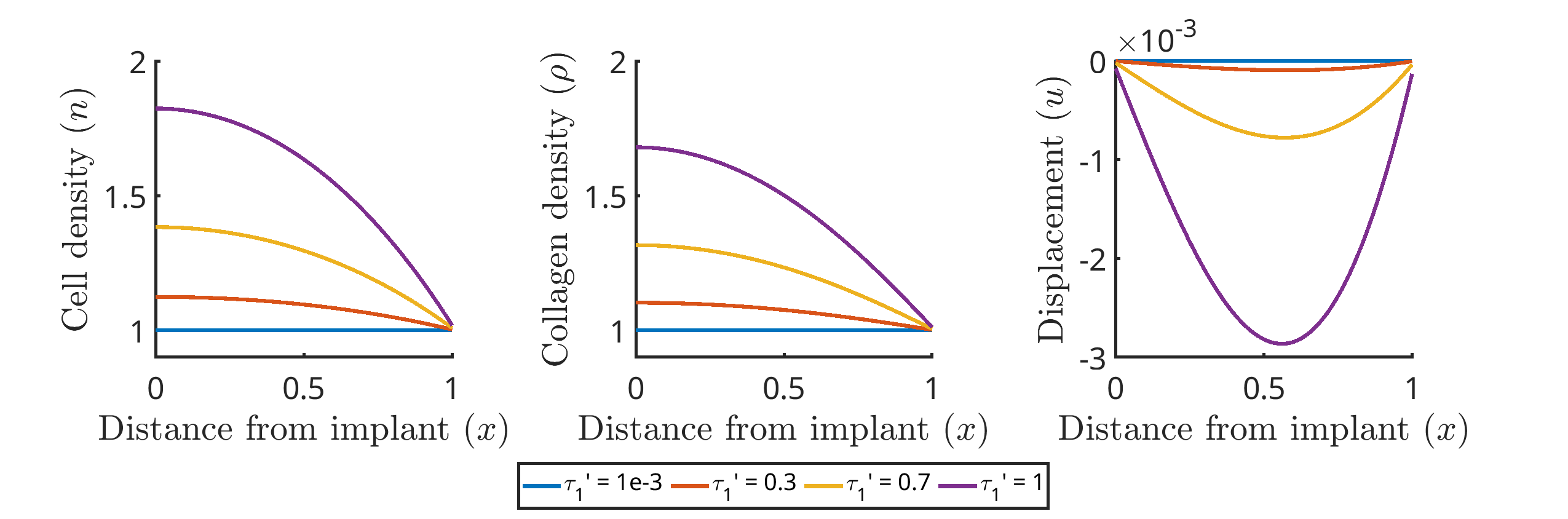}
    \caption{\textbf{More severe pathological outcome with increased traction parameter in model v-LL.} Steady state profile of cell and collagen concentration in model v-LL. We see that an elevated cell concentration is always formed at the implant site, and higher values of the dimensionless parameter $\tau'_1$ lead to higher cell and collagen concentration. 
    Initial conditions and parameter values in \eqref{eqn: ICs-params-v11}, with $\tau'_1$ values in the figure legend.}
    \label{fig:v11}
\end{figure}

\begin{figure}[H]
    \centering
    \includegraphics[width=\textwidth]{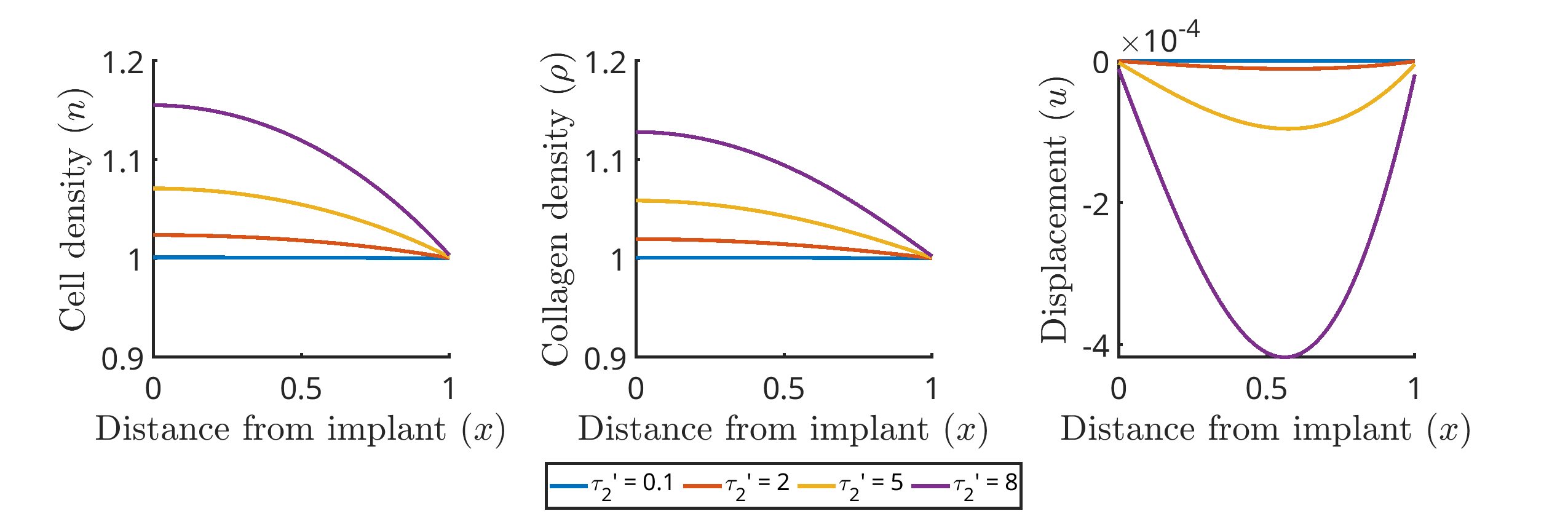}
    \caption{\textbf{More severe pathological outcome with increased traction parameter in model v-LH.} As in Figure~\ref{fig:v11}, but for model v-LH. We see that an elevated cell concentration is always formed at the implant site, and higher values of the dimensionless traction parameter $\tau'_2$ lead to higher cell and collagen concentration. Initial conditions and parameter values 
    \eqref{eqn: ICs-params-v12}, with $\tau'_2$ values in the figure legend.}
    \label{fig:v12}
\end{figure}

In Figure \ref{fig:v11}, we observe that higher values of $\tau'_1$ lead to higher steady state concentrations of cells and collagen near the implant site.
Model v-LH (stress-related cell recruitment proportional to stress, traction force a Hill function of cell density) leads to the same qualitative result as v-LL, with low ICs still leading to pathological outcomes. In Figure \ref{fig:v12}, we observe that higher $\tau'_2$ leads to higher steady state concentrations. 

\begin{figure}[H]
    \centering
    \includegraphics[width=\textwidth]{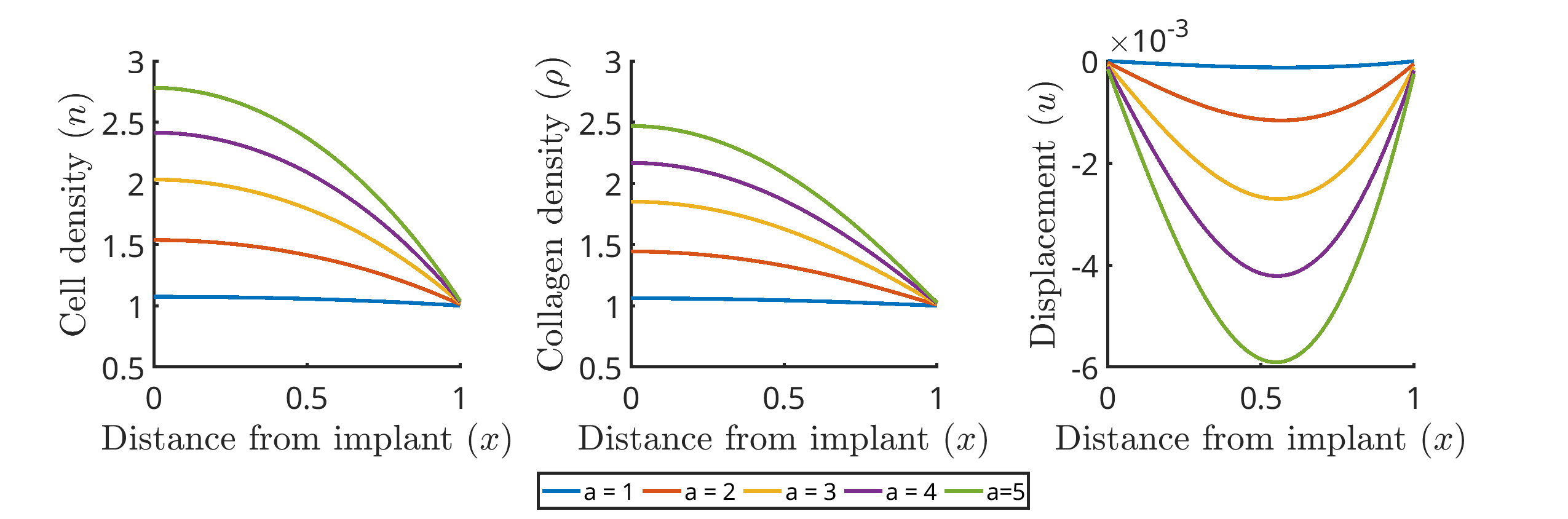}
    \caption{\textbf{More severe pathological outcome with increased recruitment parameter in model v-HL.} Steady state profile of cell and collagen concentration in model v-HL, showing the effect of the stress-induced cell recruitment amplification factor $a$. We see that an elevated cell concentration is always formed at the implant site, and higher values of $a$ lead to higher cell and collagen densities. Initial conditions and parameter values in \eqref{eqn: ICs-params-v21}, with $a$ values in the figure legend.}
    \label{fig:v21}
\end{figure}

Model v-HL (stress-related cell recruitment a Hill function of stress, traction force proportional to cell density) also leads to only pathological steady state results. Changing the values of $\tau'_1$ and the initial conditions has no effect on the final steady state profile. However, as shown in Figure \ref{fig:v21}, higher values of $a$ will lead to a higher final cell and collagen concentration at the implant site. This indicates that an elevated stress-related cell recruitment during an immune response will increase the severity of outcomes. 

\subsubsection{Bistable behavior observed with multiple cooperative feedback mechanisms}

With model v-HH (stress-related cell recruitment described by a Hill function of stress, traction force given by a Hill function w.r.t. cell density), we can classify patients into three distinct categories. (1) \textbf{Healthy patients:} these are simulated with a set of parameters that will only lead to a spatially uniform SS profile; (2) \textbf{CC patients:} these are simulated with a set of parameters that will only lead to elevated cell and collagen density profiles near the implant site; (3) \textbf{Susceptible patients:} these are bistable cases where the result depends on ICs. This is the first model where we observe bistable behavior, likely due to the incorporation of multiple Hill function assumptions.

\begin{figure}[H]
    \centering
    \includegraphics[width=0.7\textwidth]{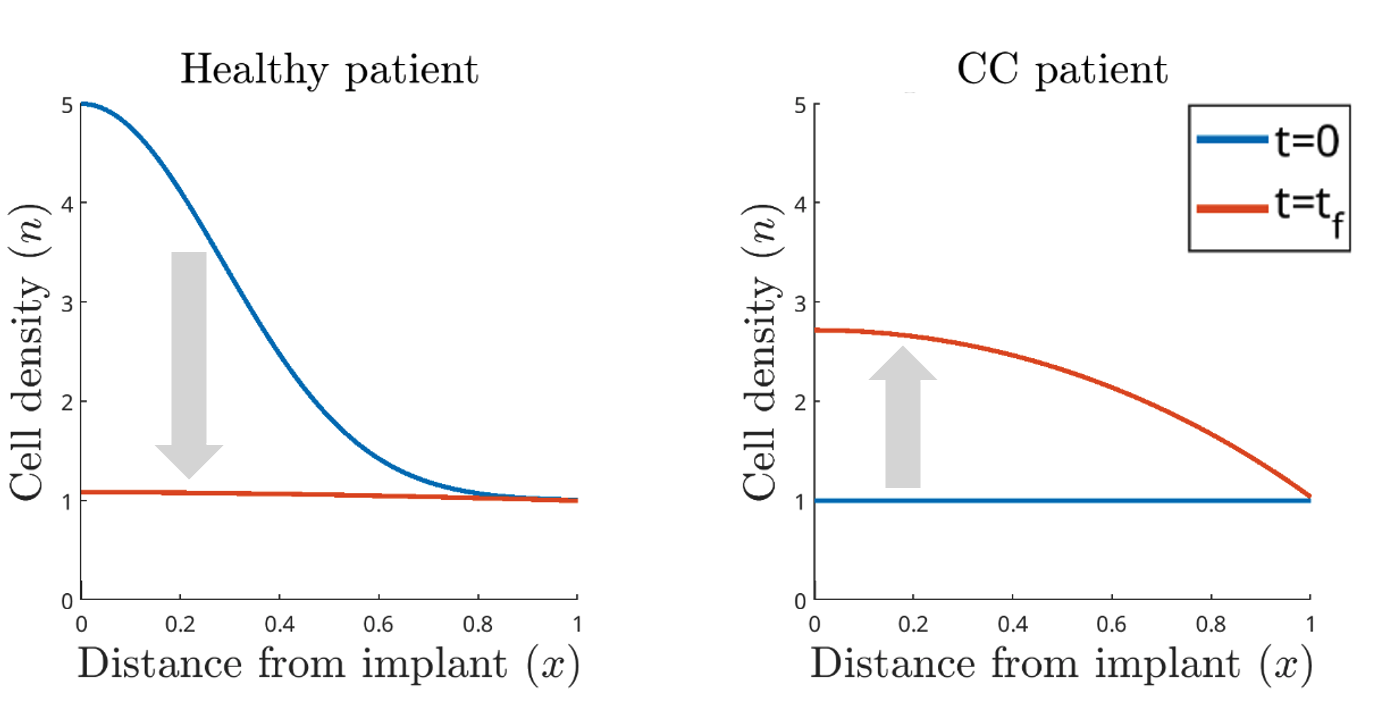}
    \caption{\textbf{Cell traction and recruitment parameters affect outcome in model v-HH.} Initial conditions (blue) and predicted final cell densities (red) in model v-HH for patients with distinct physiological parameters. Left: patient type (1) with low values of traction and recruitment parameters, $\tau'_2 = 1, a = 1$ recovers to a uniformly low ``normal'' cell density (with uniform collagen and no tissue deformation, not shown); right: patient type (2) with high values, $\tau'_2 = 20, a = 5$ develops elevated cell density, and collagen and displacement profiles (not shown) corresponding to CC. Other parameters in  \eqref{eqn: params-v22}.}
    \label{fig:v22}
\end{figure}

Figure \ref{fig:v22} shows two side-by-side panels of initial and steady state cell density for a healthy patient and a CC patient. A healthy patient will always present with healthy outcome, despite the severity of their initial immune cell recruitment, while a CC patient will always develop CC, due to their highly elevated immune response. 

\begin{figure}[H]
    \centering
    \includegraphics[width=0.8\textwidth]{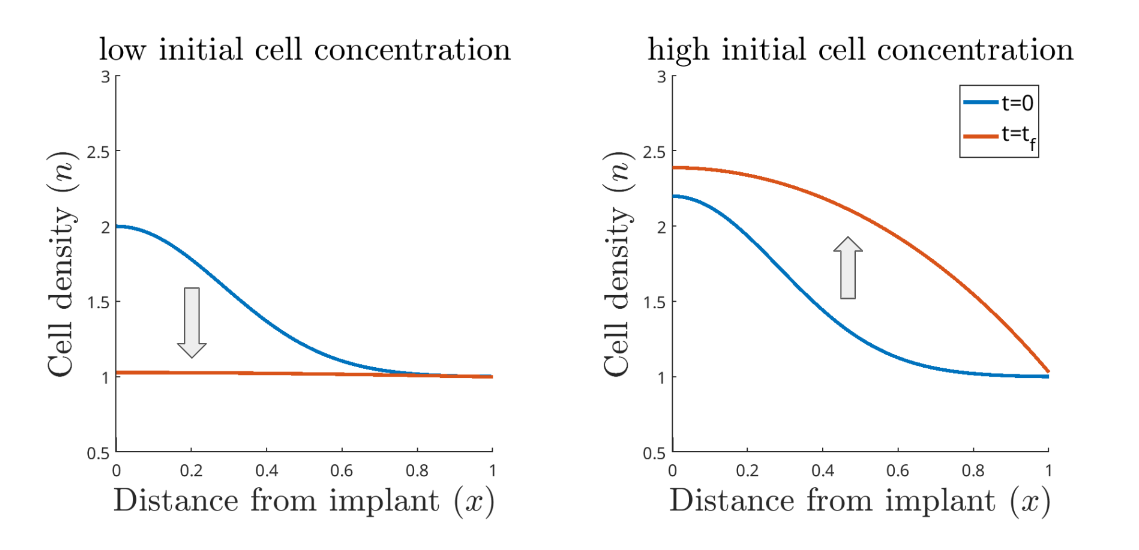}
    \caption{\textbf{Both normal or CC  outcomes are possible in the same patient according to model v-HH.} 
    Initial (blue) and final (red) cell density profiles in model v-HH with two different initial conditions representing low (left) and high (right) initial inflammation. For low initial inflammation (low initial cell density, left) the system resolves to a ``normal'' state with uniformly low cell density (as well as low collagen, and no tissue displacement not shown). For high initial inflammation (high cell level, right) model v-HH predicts CC (final steady state with elevated cell density (as well as elevated collagen and tissue displacement, not shown). This dual prediction is termed bistability. 
    Initial conditions and parameter values in \eqref{eqn: ICs-params-v22}.}
    \label{fig:v22-bistable}
\end{figure}

\begin{figure}[H]
    \centering
    \includegraphics[width=0.7\textwidth]{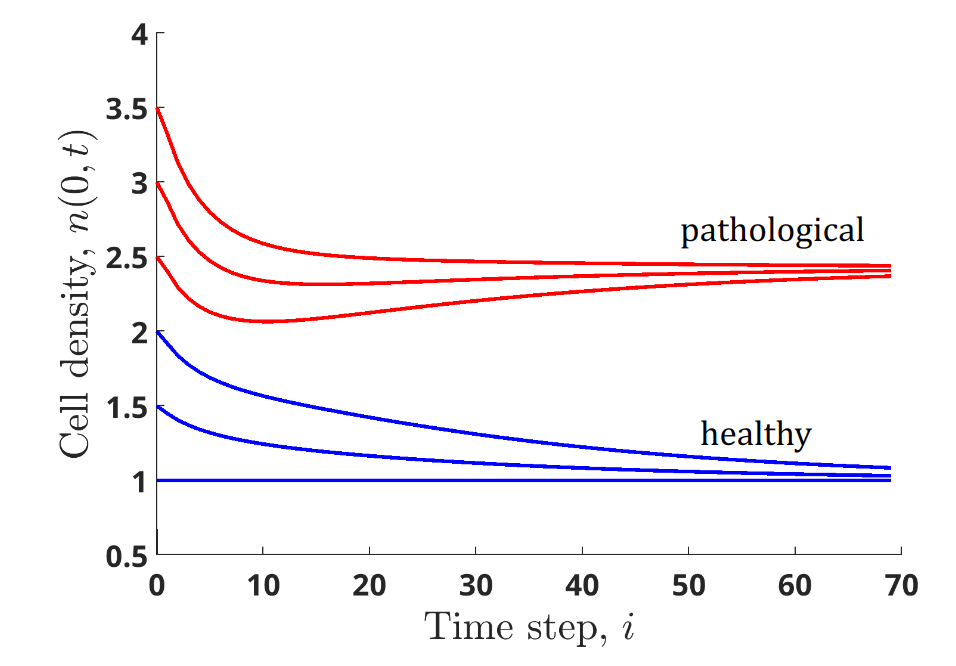}
    \caption{\textbf{Outcome depends on initial conditions in model v-HH.} Time progression of cell density at the implant site, $n(0,t)$,  predicted by  model v-HH. Blue: low initial cell densities eventually resolve to a healthy steady state value. Red: higher initial cell levels progress to a pathological CC outcome. The switching between outcomes depends on model parameters in this bistable model variant. Parameters in \eqref{eqn:params-v22-time-progression}.}
    \label{fig:v22-time-progression}
\end{figure}

Importantly, susceptible patients may have a healthy or CC outcome depending on the initial immune insult. We here characterize a pathological CC outcome with
\[
    n(0,t_{\text{end}}) > 1.1n_{SS},
\]
as immune cell density at the implant site is significantly elevated compared to the steady state value. 
In Figure \ref{fig:v22-bistable} and Figure \ref{fig:v22-time-progression}, the same set of parameters is used to generate outcomes for a single patient, experiencing distinct levels of initial immune insult or inflammation. Our results show that bistability can be observed in a certain parameter regime. 

\begin{figure}[H] 
    \centering
    \includegraphics[width=0.8\textwidth]{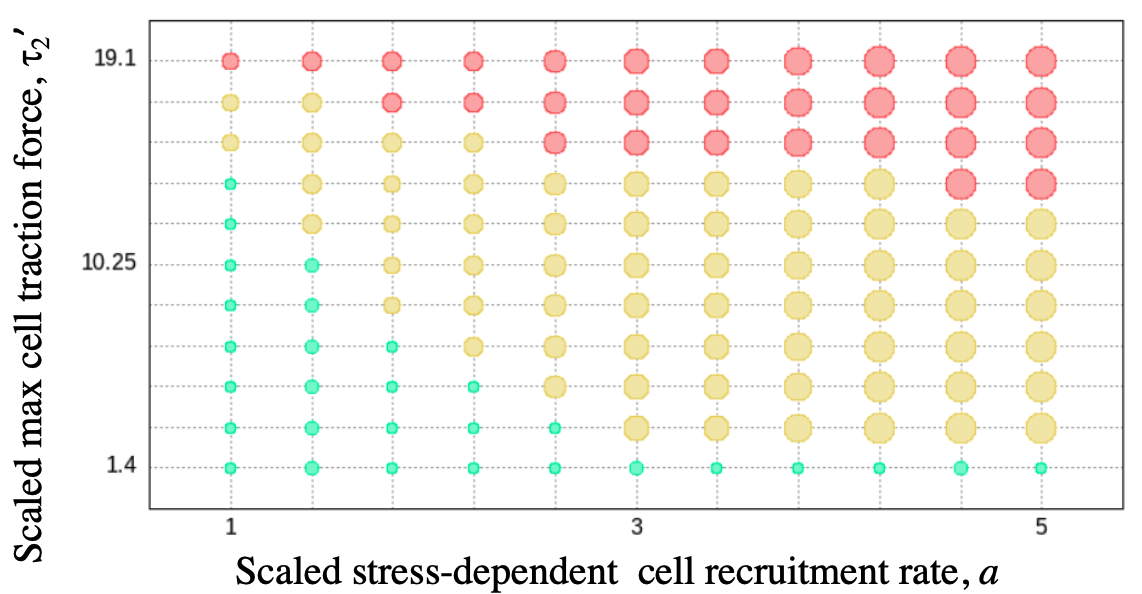}
    \caption{\textbf{Risk and severity of CC is associated with two model parameters:} Model v-HH simulations predict healthy outcomes (green), bistable behavior (yellow) and CC pathology (red), in this two-parameter plot with varying values of the dimensionless cell recruitment, $a$, and cell traction parameter $\tau_2'$. The size of each circle is proportional to the final cell density at the implant site, and represents severity. As both parameters increase, patients become more susceptible to CC with more severe outcomes.}
    \label{fig:param-plane}
\end{figure}

The steady state predicted by model v-HH is governed by the scaled cell recruitment and traction parameters, $a$ and $\tau'_2$. Figure \ref{fig:param-plane} shows how varying these parameters leads to healthy, bistable, or CC outcomes. It is evident that larger values of $a$ and $\tau'_2$ increase the risk of CC. 
Aside from the model simulations, we can also appreciate the roles of parameters from the way they appear in the steady state equations, \eqref{eqn:BVP_equations} and \eqref{eq:approxn(x)}. From the latter equation for the approximate cell density, we directly see that increasing the dimensionless stress recruitment parameter $a$ will make the elevated cell density more pronounced. Examining \eqref{eqn:BVP_equations}, we see that decreasing the collagen turnover rate $d_2$ will increase collagen density. As the tissue stress is always spatially constant ($\sigma_x=0$) at steady state, we see from \eqref{eqn:BVP_equations} (b) that larger cell-collagen traction forces (obtained for large $\tau_2'$ or dense collagen and/or high cell crowding $n\gg n_0$) can only be satisfied if there is a large values of $E_1 u_x$, which implies large gradients in tissue displacement (large $u_x$) and/or significant tissue stiffness (large $E_1$). This provides additional insights into the interwoven roles of parameters in the model.

\section{Discussion}
\label{sec: discussion}

Some breast-cancer patients are susceptible to capsular contracture following implant-based post-mastectomy breast reconstruction. This means that whether they develop CC or not depends on external factors that influence initial immune cell presence (``initial conditions''), setting the stage for interactions that our model describes. In order to account for this clinical observation, any model must be able to produce bistable behavior in some reasonable range of parameters. In such models, a low normal state can coexist with a pathological state, and initial conditions determine which of the two states will ``win''. As shown in this paper, based on the requirement for bistability, we could rule out some simple model variants that only generate a single outcome (a spatial uniform steady state). Variants v-NL and v-NH that lacked the stress-dependent recruitment term were ruled out, implying that stress related cell recruitment is an important factor in our current model. 

The requirement for bistability also restricts the type of nonlinear feedback responses. For example, if traction force is simply proportional to the cell density, or if cell recruitment is proportional to tissue stress, the model is not bistable. 

Taken together, we have identified the key ingredients leading to bistability, namely: (a) cell cooperativity (b) switch-like response to stress and (c) saturation in both responses to limit their maximal effect. 

These responses were modeled by Hill functions of cell density and of stress, respectively. Model v-HH that has all three of the above properties produces such bistable results. We believe that model variant v-HH can reasonably represents different levels of susceptibility to CC for breast implant patients. This model has parameter ranges that include the three classes of patients: those with no risk, with very high risk, and with medium risk that depends on their initial state or later immune insult. Evidence for the assumption (b) includes pathological fibrosis, where switch-like effects are common in relation to mechanical stress-mediated cell differentiation \cite{li2009spatial}. 

As demonstrated by the results in Fig.~\ref{fig:param-plane}, modeling outcomes can suggest patient risk factors that influence the likelihood and severity of CC. First, larger values of the stress-dependent cell recruitment rate $a_1$ (relative to basal rate $a_n$) increase both risk and severity. This is seen from the horizontal trend in the figure, where predominantly greater susceptibility (yellow) and severity (size) of circles occur with increasing values of $a$ (the ratio of the above rates, i.e. $a=a_1/a_n$).

From the same result, we found that increasing the maximal cell traction force $\tau_2$, which is equivalent to increasing $\tau'_2$ in Fig.~\ref{fig:param-plane}, also leads to greater risk for CC: healthy outcomes (green) are replaced by bistable (yellow) and then inevitable CC (red) in this vertical progression. This implies that the increased cell contractility and pulling forces leads to higher risks of CC. This progression could correlate with higher incidence of myofibroblasts expressing $\alpha$-smooth muscle actin or by stronger adhesion of contractile cells to the collagenous matrix, aspects not explicitly considered in the current model. 

Many other parameters can be seen to increase risk and/or severity of the capsular contracture. In hindsight, these can be understood from the core interactions and positive feedback loop of Figure~\ref{fig:model-schematic}. We see that there is a chain of events operating in the system. Cell recruitment increases local cell density, contributing to collagen buildup and a switch to the high-contractility cell state. This creates the tissue stress that, past a threshold, significantly amplifies cell recruitment. Increasing factors that contribute to that positive feedback would tend to increase and exacerbate the risk. Aside from the maximal cell traction force $\tau_2$, these include collagen production rate $b_n$, for example. Increasing the sensitivity of the
responses (by decreasing the switching-points $\sigma_0$ and $n_0$) also elevates the risk, as the vicious cycle kicks in at lower cell densities and stress. In the opposite direction, increasing the turnover of cells or collagen ($d_n, d_\rho$, respectively) is protective, as it reduces the factors that contribute to that cycle.

The model is minimal, and its limitations should be recognized. As a notable compromise between model complexity (many equations with many unknown parameters) and tractability, we have erred on the side of minimal core interactions, and a single cell type that produces and responds to stress. While this limits the direct biological application of the results, and hence of clinical relevance, it helps in analysis of the interactions of key parameters. A future extension of the model, given this initial insight, may include distinct cell populations, such as macrophages (M1 versus M2), fibroblasts, and myofibroblasts, as they appear at different stages of capsule formation and have distinct roles \cite{sadtler2016design}.

Our model treats cells as a population with no heterogeneity and ignores stochastic effects, typical of most continuum PDE models. The estimation of parameters, though done by a rigorous search of the literature is still a limitation of the work. Data specific to patient tissue samples (currently in a planning stage) or obtained from relevant targeted experimental studies for related pathologies are desirable. Values of parameters related to tissue stress and cell responses ($\sigma_0$, $n_0$, $\tau_2$, etc.) still need to be refined. 

Here we focused on BCs \eqref{eqn:bcs}. Implementing free or Neumann boundary conditions for displacement at $x=L$ gives distinct results (not shown). It would be of interest in the future to explore these differences.

While we took parameters to be spatially constant, in actual fact, they may depend on the environment in proximity (or far from) the implant, and they may vary as the process of healing (or pathology) evolves with time. There is evidence that cell traction forces increase with stiffness of the cell's environment \cite{trichet2012evidence}. Hence the parameters $\tau_i$ that represent the magnitudes of those traction forces (here taken as constant) may increase as the tissue stiffens in early stages of CC, adding important nonlinear effects to consider.

Tissue properties also depend on cell density. Van Oosten et al. \cite{van2019emergence} showed that a tissue composed of cells embedded in ECM becomes stiffer when it is compressed (essentially due to the fact that cells are volume-conserving). This effect is cell-density dependent, so as cells are recruited to the implant, they also affect the material properties of the tissue. Such nonlinear tissue rheology implies that the stiffness parameter $E$ (assumed constant in our model) changes as strain and cell density changes. It also suggests that including such nonlinear rheology tissue properties could be important to take into account in future work. 

The 1D geometry of the domain also restricts the applicability of the domain. Indeed, to classify cases into ``normal", at medium risk, or at high risk of CC, our spatially 1-dimensional model suffices, and considerably simplifies the solid-mechanics treatment. This is a geometric simplification that could be improved by considering 2D and 3D anisotropic tissue models, as in \cite{villa2021mechanical,ben2015morpho}. The alignment of collagen fibers would then affect anisotropic tissue properties that would be require suitable constitutive laws. More realistic geometry in 3D can be accommodated but would required finite-element simulations. However our goal here was to identify the key elements that are necessary for CC. We are now in a position to port these elements to more realistic settings.

Despite such model limitations, overall conclusions do not depend on specific parameter values, but rather on the structure of the model and the types of feedback assumed between stress and cell recruitment. 

Overall, beneficial treatment strategies to reduce CC risk can be suggested, even in this minimal model investigation. Briefly, treatments that prevent initial inflammation (anti-inflammatory agents) can protect a patient against the dangerous ``initial conditions'' that lead to CC. Local regulation of cell contractility at the implant site could decrease the parameter $\tau_2$ whose high values correlate with high risk (Figure~\ref{fig:param-plane}). Inhibitors of Rho/ROCK that suppress myosin contraction have such properties. Drugs that locally regulate collagen deposition could be beneficial (Smad 2/3/4). Local inhibitors of TGF-$\beta$ (whose downstream effects include both ROCK and Smad upregulation) may also be relevant \cite{jiang2015estradiol}. Future models that include more details of such underlying biochemical signalling could be useful. While current literature identifies the over-expression of TGF-$\beta$ as an important indicator of abnormal wound healing, its role in CC is yet to be explored in experimental and modeling studies.

\section*{Acknowledgements}
LEK is supported by an NSERC Discovery grant. At early stages of this project, YX was supported by a USRA and then an RA funded by NSERC funding to LEK. Funding for this project has been granted by a Cancer Research Society operating grant to KVI and LEK.

\begin{appendices}

\section{Numerical schemes for the system of PDEs}
\label{appendix: numerics}

Numerical simulation results for the time-dependent, spatially one-dimensional system of PDEs (\eqref{eq: non-dimensioned_general_model} (a)-(d)) are obtained via the method of lines. MATLAB code for all numerical schemes and figure generation is available at \url{https://github.com/Yuqi-eng/Thesis-work}.

Let $K$ be the number of grid points, we have cells, collagen and displacement approximated as

\[
    n(t,x_i) = N_i(t); \rho(t,x_i) = P_i(t); u(t,x_i) = U_i(t), \text{ for $1\leq i\leq K$.}
\]

In order to implement the boundary conditions in \eqref{eqn:bcs}, we have

\begin{subequations}
{
\[
    \begin{array}{ll}
    \text{Cells}\qquad &N_0(t)= N_2(t), N_{K+1}(t) = N_{SS}, \\
    \hfill \\
    \text{Collagen}\qquad &P_0(t)= P_2(t), P_{K+1}(t) = P_{SS}, \\
    \hfill \\
    \text{Displacement}\qquad &U_0(t) = U_{k+1}(t) = 0.
    \end{array}
\]
} 
\label{eqn:bcs_implementation}
\end{subequations}

For the sake of convenient notations, we also approximate the stress related recruitment term and traction forces as

\[
    f(\sigma(t,x_i)) = F_i(t), g(n(t,x_i),\rho(t,x_i)) = G_i(t), \text{ for $1\leq i\leq K$.}
\]

And we let 

\[
    G_0(t) = G_2(t), G_{K+1}(t) = g(N_{SS}, P_{SS}).
\]

We use the following $K\times (K+2)$ matrices to compute first and second derivatives using second order centered finite differencing.

$M_x = \frac{1}{2\Delta x}\begin{bmatrix}
-1 & 0 & 1 & 0 & 0 & 0 & \cdots & 0\\
0 & -1 & 0 & 1 & 0 & 0 & \cdots & 0\\
&&&\cdots&&& \\
0 & \cdots & 0 & 0 & 0 & -1 & 0 & 1\\
\end{bmatrix}$, $M_{xx} = \frac{1}{\Delta x^2}\begin{bmatrix}
1 & -2 & 1 & 0 & 0 & 0 & \cdots & 0\\
0 & 1 & -2 & 1 & 0 & 0 & \cdots & 0\\
&&&\cdots&&& \\
0 & \cdots & 0 & 0 & 0 & 1 & -2 & 1\\
\end{bmatrix}$.

Then let $\Tilde{N} = \begin{bmatrix}
   N_0 \\
   N_1 \\
   \vdots \\
   N_K \\
   N_{K+1} \\
\end{bmatrix}$, $\Tilde{P} = \begin{bmatrix}
   P_0 \\
   P_1 \\
   \vdots \\
   P_K \\
   P_{K+1} \\
\end{bmatrix}$, $\Tilde{U} = \begin{bmatrix}
   U_0 \\
   U_1 \\
   \vdots \\
   U_K \\
   U_{K+1} \\
\end{bmatrix}$, $\Tilde{G} = \begin{bmatrix}
   G_0 \\
   G_1 \\
   \vdots \\
   G_K \\
   G_{K+1} \\
\end{bmatrix}$, we get 

\[
    \frac{\partial y(t,x_i)}{\partial x} = M_x\Tilde{Y}, \qquad \frac{\partial^2 y(t,x_i)}{\partial x^2} = M_{xx}\Tilde{Y}, 
\]
where 
\[
    y = \{n, \rho, u, g\}, \qquad \Tilde{Y} = \{\Tilde{N}, \Tilde{P}, \Tilde{U}, \Tilde{G}\}. 
\]

Since $U_1 = U_K = 0$, we can obtain the second derivative of the velocity vector wrt. space with

\[
    \frac{\partial^2 v}{\partial x^2} = M_{xx}\Tilde{U}' = M_{xx}\begin{bmatrix}
        0 \\
        U'_1\\
        \vdots\\
        U'_K\\
        0
    \end{bmatrix}.
\]

We use an upwinding scheme to compute the gradient of deformation flux for cells and collagen. First, we compute the flux at each grid cell
\[
    [\mathscr{F}(U'(t), Y(t))]_i = \begin{cases} 
      U_i'(t)Y_{i-1}(t) & U_i'(t) > 0, \\
      U_i'(t)Y_i(t) & U_i'(t) \leq 0,
   \end{cases}
\]

where $Y(t) = N(t)$ or $P(t)$.

Then, we obtain the gradient of this flux as a $K\times1$ column vector 

\[
\nabla\mathscr{F} = \begin{bmatrix}
   \frac{1}{\Delta x}(\mathscr{F}_2-\mathscr{F}_1) \\
   (M_x\mathscr{F})_2 \\
   \vdots \\
   (M_x\mathscr{F})_{K-1} \\
   \frac{1}{\Delta x}(\mathscr{F}_{K}-\mathscr{F}_{K-1}) \\
\end{bmatrix},
\]

such that 

\[
    \frac{\partial (nv)}{\partial x} = \nabla\mathscr{F}(U',N), \qquad \frac{\partial (\rho v)}{\partial x} = \nabla\mathscr{F}(U',P).
\]

Finally, we solve the following time dependent system of ODEs using the fully implicit MATLAB solver ode15i.  
\begin{subequations}
    \begin{align}
    \text{Cells}\qquad      &N_i' - (M_{xx}\Tilde{N})_i + [\nabla\mathscr{F}(U', N)]_i - [\mathscr{K}_n(N, P)]_i = 0,  \\
    \text{Collagen}\qquad    &P_i' - D_2(M_{xx}\Tilde{P})_i + [\nabla\mathscr{F}(U', P)]_i - [\mathscr{K}_{\rho}(N, P)]_i = 0,\\ \text{Force balance}\qquad      &E_1(M_{xx}\Tilde{U})_i + \eta_1(M_{xx}\Tilde{U'})_i + (M_x\Tilde{G})_i = 0,
    \end{align}
    \label{eqn:numerical_ODE}
\end{subequations}
where
\[
    \mathscr{K}_n(N, P) = 1 + F(t) - N(t), \qquad \mathscr{K}_{\rho} = N(t) - d_2P(t)
\] includes terms without spatial or temporal derivatives. \eqref{eqn:numerical_ODE} (c) results from combining \eqref{eq: non-dimensioned_general_model} (c)-(d).

\section{Numerical simulation of the BVP}
\label{appendix: BVP}
Using a MATLAB BVP solver, we can approximate the steady state solution of the system described in \eqref{eqn:BVP_equations} (a)-(d). In order to do this, we rewrite the equations into the following form. 
\begin{subequations}
    \begin{align*}
        n_{xx} &= n - 1 - a\frac{\sigma^{k_1}}{1+\sigma^{k_1}},\\
        \rho_{xx} &= \frac{1}{D_2}(d_2\rho - n),\\
        u_x &= \frac{1}{E_1} (\sigma-\tau'_2 \frac{n^{k_2}}{N_0^{k_2} + n^{k_2}}\rho),\\
        \sigma_x &= 0.\\
    \end{align*}
\end{subequations}    
We then obtain the vector inputs as required for MATLAB's BVP solver, see details in MATLAB documentation and the provided code. 

\section{Parameter estimation methods}
\label{appendix: ParamDetails}

\subsection{Summary Table}
Table \ref{tab:paramsMeaning} is a summary of all parameters in \eqref{eqn:full_model}. This section details literature support for estimating some of these values.

\begin{table}[H]
    \centering
    \renewcommand{\arraystretch}{1.2}
    \begin{tabularx}{0.92\textwidth}{|c|X|c|} \hline
        Parameter & Physical Meaning & Unit \\ \hline
        
        $D_n$ & Rate of random cell motility & $\mu m^2/s$ \\ \hline
        
        $a_n$ & Basal cell recruitment rate & cells/$(\mu m^3\cdot s)$ \\ \hline

        $a_0$ & Cell density recruitment per unit stress & cells/($\mu m^3\cdot s\cdot Pa$) \\ \hline

        $a_1$ & Maximal stress-related cell recruitment rate & cells/$(\mu m^3\cdot s)$ \\ \hline

        $\sigma_0$ & Level of stress for elevated rate of cell recruitment & $Pa$ \\ \hline

        $k_1$ & Hill function coefficient for stress related cell recruitment 
        & 1 \\ \hline

        $d_n$ & Cell turnover rate & $/s$ \\ \hline

        $D_{\rho}$ & Rate of random collagen reorganization & $\mu m^2/s$ \\ \hline

        $b_n$ & Collagen production rate per cell & $mg/(\text{cells}\cdot s)$ \\ \hline
        
        $d_{\rho}$ & 
        Collagen decay rate & $/s$ \\ \hline

        $E$ & Elastic modulus of the tissue & $Pa$ \\ \hline

        $\eta$ & Viscosity coefficient of the tissue & $Pa\cdot s$ \\ \hline

        $\tau_1$ & Coefficient of cell traction & Pa/(cells/$\mu$m$^3$)/(mg/$\mu$m$^3$) \\ \hline

        $\tau_2$ & Coefficient of cell traction & Pa/(mg/$\mu$m$^3$) \\ \hline

        $n_0$ & Critical cell density for  traction force cooperativity
        & cells/$\mu m^3$ \\ \hline

        $k_2$ & Hill function coefficient for cell traction & 1 \\ \hline      
    \end{tabularx}
    \caption{Parameters used in \eqref{eqn:full_model} and their units.}
    \label{tab:paramsMeaning}
\end{table}

\subsection{Conversion factors}

Some parameters are given in units of mL and days. We translate all values to time and size scale of cells, namely $\mu$m and $s$.
\[
1 \text{mL} = 10^{12}\mu \text{m}^3.
\]
\[
1 \, \text{day} = 8.6\times 10^{4} s \approx 10^5 s.
\]
\textbf{Density units}: We find a variety of units in the literature. For example, in \cite{wigginton2001model}, density is given as cells/ml. To convert to cells/$\mu$m$^3$, we use
\[
1 \text{cell/ml}= 10^{-12} \text{cells}/\mu\text{m}^3.
\]

\begin{figure}[ht]
    \centering
    \includegraphics[width=0.3\textwidth]{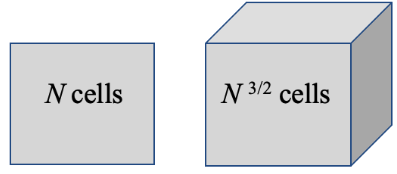}
    \caption{Based on samples viewed in 2D microscopy slides, researchers often estimate the number of cells per unit area. We here assume that this reflects a view if what is an underlying uniform distribution of cells in the tissue volume. To convert to density per unit volume we make the simple assumption: if $N$ cells are observed per unit area in a thin 2D section, then it implies there are $N^{3/2}$ per unit volume.}
    \label{fig:AreaToVolume}
\end{figure}

\textbf{Density in 2D vs 3D}: Many papers provide cell density per unit area in microscopy slides. To convert this to 3D  we make a simple assumption of uniform density throughout a tissue. (See Fig.~\ref{fig:AreaToVolume}.) Hence if we observe $N$ cells per unit area, then it implies that the cell inside the whole volume would be approximated as 
\[
N^{3/2} \text{cells per unit volume}.
\]

\subsection{Literature Support}
Since our population of ``cells'' is a composite of various cell types, we used information about cell types implicated in CC, under comparable contexts, to estimate appropriate parameter ranges.
Data for macrophage density in normal tissue, the macrophage death rate, and recruitment rate in normal tissue and in tuberculosis (TB) is available from the Kirschner lab, e.g.,   \cite{wigginton2001model}. Here, the authors estimate normal macrophage density as $10^5$ cells/mL, which equates to our value $n_{SS}\approx 10^{-7}$ cells/$\mu$m$^3$. They use these values to estimate the basal macrophage recruitment rate to be around 330-440 cells /(mL $\cdot$ day), which is equivalent to 3.82-5.1$\times 10^{-15}$ cells/($\mu$m$^3\cdot$ s).  Macrophage-induced myofibroblast recruitment rate can be up to 2-10 times the basal recruitment of myofibroblasts in uninjured tissues \cite{katz1991human}. Therefore, we estimate $a_n \approx$ 0.76-5.2$\times 10^{-14}$cells/($\mu$m$^3\cdot$ s).

The time-course of monocyte/macrophage populations over a few days of wound-healing (right after a skin injury) is quantified in \cite{agaiby1999immuno} Table 2 and Figure 2, based on data from rats. The observed cell count in 2D sections increases from low levels (around 2) to high levels (around 40), i.e. a 20 fold increase, over a period of 1-3 days. If we correct this to account for the cells in a volume, then the increase is $\approx 20^{3/2}\approx 90$. From the rate of decay they measured, we can also estimate a half-life of approximately 10 days, which implies a decay rate of $\approx 0.1$/day = $10^{-6}$/s. In the models of non small cell lung cancer (NSCLC), \cite{bartha2022mathematical} suggests a turnover rate of 0.1-1$\times 10^{-5}$/s for macrophages. In a more recent round of papers, with a focus on the effect of cytokines such as TNF on macrophage recruitment, macrophage death rates are given as 4-6 $\times 10^{-3}$/day, which is in the range of 4-6 $\times 10^{-8}$/s (Denise Kirschner, table of parameters, personal communication Oct 23, 2023). Once an immune response is triggered, there are cytokines that inactivate macrophages, which tends to increase the turnover rates (i.e. rates of inactivation of activated macrophages) by a factor of 10-20. This brings the range of turnover rate to 0.4-1.2 $\times 10^{-6}$/s. According to \cite{baum2011fibroblasts} which cites \cite{frangogiannis2000myofibroblasts}, myofibroblasts peak 1-2 weeks after cardiac injury, and decay 21-28 days afterwards. This suggests a turnover rate of $\approx 1/28=0.035$/day $\approx 3.5 \times 10^{-7}$/s. From all the sources above, we obtain $d_n \approx 0.35$-$10 \times 10^{-6}$/s.

Typical collagen density in mice mammary tissues is  in the range of $ 1.3$-$3\times 10^{-12}$mg$/\mu$m$^3$  according to \cite{provenzano2008collagen}.
In \cite{illsley2000increased}, collagen production by control group dermal fibroblasts is 29.8-82.84 pmoles hydroxyproline (OHpro)/$10^5$ cells/hour. Hyproxyproline accounts for 15-22\% of amino acid in collagen, with collagen having a molar mass of $3\times 10^5$ g/mol. We estimate a normal rate of collagen production by fibroblasts at $b_n \approx 0.5$ - $1.4\times10^{-11}$mg/cells/s. 
In \cite{cleutjens1995regulation}, we get some information about the timescale of collagen turnover in the context of cardiac injury (in rats). There are various collagen types that are more (Types I) or less (Type  III) resistant to degradation. Collagen breakdown starts hours after injury and continues for 3-4 days. According to \cite{cleutjens1995regulation}, collagenase activity is already increased by day 2 (by 4.5-fold) and is highest at day 7 (6.5-fold). This suggests that we can use a collagen decay timescale of roughly 1 week in the context of healing, so $d_{\rho} \approx 1.6 \times 10^{-6}$/s.

The elastic modulus for normal breast tissues found in literature ranges from $E \approx $ 2.34-4.16 kPa, while in fibrocystic and cancerous breast tissues it can increase to roughly 5-10 times its normal value.\cite{samani2007elastic}.
In phantom tissues, $\eta/E$ is approximately 7 ms \cite{liu2023real}. We can use this to estimate $\eta \approx $ 16.4-29.1 N$\cdot$ s/m$^2$.

In \cite{chen2007alpha}, individual cell traction forces were detected at 100-300 $Pa$ for fibroblasts and myofibroblasts, on a collagen gel with density 100$\mu$g/mL. If we assume that at each point in our 1D domain, traction force is generated by 1 cell, then $\tau_1 n \approx \tau_2 f(n) \approx$ 1-3$\times 10^{15}\text{Pa}/(\text{mg}/\mu\text{m}^3)$. From here we get the estimates $\tau_1 \approx$ 1-3$\times 10^{22}\text{Pa}/(\text{cells}/\mu\text{m}^3)/(\text{mg}/\mu\text{m}^3)$ and $\tau_2 \approx$ 1-3$\times 10^{15}\text{Pa}/(\text{mg}/\mu\text{m}^3)$.

\section{Dimensionless parameters}
\label{sec:AppdxDim'lessPar}

\begin{table}[H]
    \centering
    \renewcommand{\arraystretch}{1.2}
    \begin{tabularx}{0.6\textwidth}{|c|X|} \hline
        Parameter & Value  \\ \hline

        $n_{SS}^* = n_{SS}a_n/d_n$ & 1 \\ \hline

        $\rho_{SS}^* = \rho_{SS}d_n^2/a_nb_n$ & 1 \\ \hline
        
        $a = a_1/a_n$ & 1-5  \\ \hline

        $D_2 = D_{\rho}/D_n$ & 0.1 \\ \hline

        $d_2 = d_{\rho}/d_n$ & 1 \\ \hline

        $N_0 = n_0d_n/a_n$ & 2 \\ \hline
        
        $k_1$ & 5 \\ \hline

        $k_2$ & 5\\ \hline  

        \multirow{2}{*}{$E_1 = E/S$}
        & 1,\hfill for v-0L, v-0H \\ \cline{2-2}
        & 100,\hfill for v-LL, v-LH, v-HL, v-HH \\ \hline

        \multirow{2}{*}{$\eta_1 = \eta d_n/S$}
        & $10^{-8}$,\hfill for v-0L, v-0H \\ \cline{2-2}
        & $10^{-6}$,\hfill for v-LL, v-LH, v-HL, v-HH \\ \hline

        \multirow{2}{*}{$\tau'_1 = \tau_1a_n^2b_n/Sd_n^3$}
        & $10^{-6}$-$10^{-2}$,\hfill for v-0L, v-0H \\ \cline{2-2}
        & $10^{-4}$,\hfill for v-LL, v-LH, v-HL, v-HH \\ \hline

        \multirow{2}{*}{$\tau'_2 = \tau_2a_nb_n/Sd_n^2$}
        & $10^{-4}$-0.1,\hfill for v-0L, v-0H \\ \cline{2-2}
        & $10^{-2}$-30,\hfill for v-LL, v-LH, v-HL, v-HH \\ \hline
    \end{tabularx}
    \caption{Table of dimensionless parameters. Some can be estimated from literature data given in Appendix \ref{appendix: ParamDetails}. Other parameters are a range of estimated fold-multiples of the basal rates that stem from the effect of injury, surgery, immune response etc. }
    \label{tab:dimensionlessParams}
\end{table}

Many of these estimates are based on modifications of the preliminary parameter estimates that are also suitable for stability of the numerical simulation. In what follows, we use the notation * to denote dimensionless quantities. In the text of the paper, we later drop the *'s for notational convenience.

\textbf{Homogeneous steady state cell density,} $\mathbf{n_{SS}}$: A typical steady state cell density is $n_{SS} \approx 10^{-7}$ cells/$\mu$m$^3$. We chose this as the reference cell density, that is we define
\[
N = n_{SS}=\frac{a_n}{d_n} = 10^{-7} \text{cells}/\mu \text{m}^3.
\]
(After scaling, the dimensionless HSS cell density is $n_{SS}^*=n_{SS}/N =  1$).

The above typical cell density, and other information is consistent with typical recruitment and turnover parameter values $a_n = 5\times10^{-14}$ cells/($\mu$m$^3\cdot$ s) and $d_n = 5\times10^{-7}$/s.

\textbf{Initial immune cell density at the implant site, $\mathbf{n(0,0)}$}:
According to \cite{boyce2001inflammatory}, in  keloid scars, activated macrophage density is 2.3 times its density in uninjured tissue. In chronic nerve compression injuries, the ratio of macrophage density in injured tissues to normal tissues is 10-12 times \cite{gupta2006spatiotemporal}. We therefore estimate $n(0,0) = $1-10$\times n_{SS}$. A similar factor applies to the scaled cell densities.

\textbf{Basal steady state collagen density, }$\boldsymbol{\rho}\mathbf{_{SS}}$: We use a typical collagen density, $\rho_{SS}$ = $2\times10^{-12}$mg/$\mu$m$^3$ for the basal collagen steady state. We keep the same choices of values for $a_n, b_n$, and choose $b_n = 10^{-11}$mg/cells/s. For the collagen reference value, we hence use
\[
    \varrho = \frac{a_nb_n}{d_n^2} \approx 2\times10^{-12}\text{mg}/\mu\text{m}^3, 
\]
The scaled collagen HSS is then $\rho_{SS}^*=\rho_{SS} /\varrho= 1$.

\textbf{Scaled cell recruitment rate,}
$\mathbf{a}$: 
The dimensionless parameter $a$ represents the ratio of stress-related cell recruitment rate to basal recruitment rate. There is no supporting literature on which to base specific estimates, so we assume that $1 \le a \le 5$. In this way, the effect of stress is at least to double cell recruitment (when $a=1$) or increase it significantly (for $a\ \approx 5$).

\textbf{Diffusion rates,}
$\mathbf{D_2}$: 
$D_2$ represents a ratio of random collagen rearrangement, $D_{\rho}$, to the rate of random cell motility, $D_n$. Collagen does not ``diffuse'' on the timescale of cell random motility, though it gets reorganized and randomly redistributed through cell activity. We mainly need to consider $D_{\rho}$ in the model to regularizing the numerical method. We found that $D_2 = 0.1$ suffices for that purpose.

\textbf{Turnover rate,} $\mathbf{d_2}$: 
This dimensionless parameter is the ratio of collagen to cell turnover rates. To fit the desired model behaviour (uniform steady state collagen density in healthy cases), we took $d_2 \approx 1$. This implies that at early stages, cells turnover and collagen turnover operate on a similar timescale, as cells transit in and out of the injured area and actively remodel (secrete and degrade) collagen.

\textbf{Reference stress level} $\mathbf{S}$:
Typically stress in human tissue is measured in the unit of kPa (ranges observed $\approx$ 0.1-2.4 kPa), values that guide some choices we made.

However, as there are three model cases considered, we select three appropriate choices of the dimensionless parameter $S$ as a reference scale. 

For v-LL, v-LH, we can assume that $a_0\sigma \approx$ 5-10$a_n$ for an elevated effect of stress-related cell recruitment rate compared to the basal rate. Typically stresses in human tissues is measured in the unit of kPa (ranges observes $\approx$ 0.1-2.4 kPa), therefore we can assume $S = \frac{a_n}{a_0} \approx \text{0.01-0.48}\times 10^3$ Pa. For v-HL, v-HH, we may assume that the rate of activation for stress-related cell recruitment, $\sigma_0$, falls into the same order of magnitude as $\frac{a_n}{a_0}$, due to the lack of support literature on this value.

\textbf{v-NL, v-NH}: We identify $S = E \approx 2.34\text{-}3.16\cdot 10^3$ Pa. We select 
\[
S = E \approx 3\cdot 10^3 \text{ Pa}.
\]
Then
\[
E_1 = \frac{E}{S} \approx 1, \eta_1 = \frac{\eta d_n}{S} \approx 10^{-8}, \tau'_1 = \frac{\tau_1a_n^2b_n}{Sd_n^3} \approx 10^{-6}\text{-}10^{-2}, \tau'_2 = \frac{\tau_2a_nb_n}{Sd_n^2} \approx 10^{-4}\text{-}0.1.
\]

\textbf{v-LL, v-LH, v-HL, v-HH}: For v-LL, v-LH, we assume that $a_0\sigma \approx$ 5-10$a_n$ for the stress-related amplification of cell recruitment (relative to the basal rate). Based on the typically stress in human tissue, we assume $S = a_n/a_0 \approx \text{0.01-0.48}\times 10^3$ Pa. v-HL, v-HH share these parameter values as discussed. Hence, we pick
\[
S \approx \frac{a_n}{a_0} \approx \sigma_0 \approx \text{30}\text{ Pa}.
\]
Then 
\[
E_1 \approx 100, \eta_1 \approx 10^{-6}, \tau'_1 \approx 10^{-4}\text{-}1, \tau'_2 \approx 10^{-2}\text{-}30.
\]

\textbf{Hill function coefficients:} In order to simulate a smooth switch-like mechanism, we set $k_1 = k_2 = 5$. Additionally, we set $N_0 = 2$ based on previous arguments for elevated cell recruitment.

\section{Numerical parameters used}
\label{appendix: parameters}

Initial guesses and parameters used to generate Figure \ref{fig:analytic_bvp_compare}:
\begin{subequations}
\begin{align}
    \text{low initial guess \& low IC:}\qquad &n(x) = 1, \rho(x) = 1, u(x) = 0, \sigma(x) = 0. \\
    \text{high initial guess:}\qquad &n(x) = 5, \rho(x) = 1, u(x) = 0, \sigma(x) = 5. \\
    \text{high IC:}\qquad &n(0,x) = 1+ 4e^{-(\frac{x}{0.4})^2}, \rho(0,x) = 1, u(0,x) = 0. \\
    \text{Parameters:}\qquad &D_2 = 0.1, d_2 = 1, a = 4, E_1 = 100, \eta_1 = 10^{-6}, \tau'_2 = 10.
\end{align}
\label{eqn:ICs-params-BVP}
\end{subequations}
Initial conditions (ICs) and parameters used to generate Figure \ref{fig:v01-02}:
\begin{subequations}
\begin{align}
    \text{ICs:}\qquad &n(0,x) = 1 + 4e^{-(\frac{x}{0.4})^2}, \rho(0,x) = 1, u(0,x) = 0. \\
    \text{Parameters:}\qquad &D_2 = 0.1, d_2 = 1, E_1 = 1, \eta_1 = 10^{-8}, \tau'_1 = 10^{-4}.
\end{align}
\label{eqn: ICs-params-v01-02}
\end{subequations}
ICs and parameters used to generated Figure \ref{fig:v11-time} and Figure \ref{fig:v11}:
\begin{subequations}
\begin{align}
    \text{ICs:}\qquad &n(0,x) = 1, \rho(0,x) = 1, u(0,x) = 0. \\
    \text{Parameters:}\qquad &D_2 = 0.1, d_2 = 1, E_1 = 100, \eta_1 = 10^{-6}.
\end{align}
\label{eqn: ICs-params-v11}
\end{subequations}
ICs and parameters used to generated Figure \ref{fig:v12}:
\begin{subequations}
\begin{align}
    \text{ICs:}\qquad &n(0,x) = 1, \rho(0,x) = 1, u(0,x) = 0. \\
    \text{Parameters:}\qquad &D_2 = 0.1, d_2 = 1, E_1 = 100, \eta_1 = 10^{-6}, N_0 = 2, k_2 = 5.
\end{align}
\label{eqn: ICs-params-v12}
\end{subequations}
ICs and parameters used to generated Figure \ref{fig:v21}:
\begin{subequations}
\begin{align}
    \text{ICs:}\qquad &n(0,x) = 1, \rho(0,x) = 1, u(0,x) = 0. \\
    \text{Parameters:}\qquad &D_2 = 0.1, d_2 = 1, E_1 = 100, \eta_1 = 10^{-6}, k_1 = 5, \tau_1' = 0.7.
\end{align}
\label{eqn: ICs-params-v21}
\end{subequations}
Parameters used to generate Figure \ref{fig:v22}:
\begin{subequations}
\begin{align}
    \text{Parameters:}\qquad D_2 = 0.1, d_2 = 1,  E_1 = 100, \eta_1 = 10^{-6}, k_1 = 5, k_2 = 5, N_0 = 2.
\end{align}
\label{eqn: params-v22}
\end{subequations}
ICs and parameters used to generate Figure \ref{fig:v22-bistable}:
\begin{subequations}
\begin{align}
    \text{ICs(low):}\qquad &n(0,x) = 1+ e^{-(\frac{x}{0.4})^2}, \rho(0,x) = 1, u(0,x) = 0. \\
    \text{ICs(high):}\qquad &n(0,x) = 1+ 1.2e^{-(\frac{x}{0.4})^2}, \rho(0,x) = 1, u(0,x) = 0. \\
    \text{Parameters:}\qquad &D_2 = 0.1, d_2 = 1,  E_1 = 100, \eta_1 = 10^{-6}, k_1 = 5, k_2 = 5, N_0 = 2, \tau'_2 = 5, a = 4.
\end{align}
\label{eqn: ICs-params-v22}
\end{subequations}
Parameters used to generate Figure \ref{fig:v22-time-progression}:
\begin{subequations}
\begin{align}
    \text{Parameters:}\qquad &D_2 = 0.1, d_2 = 1,  E_1 = 100, \eta_1 = 10^{-6}, k_1 = 5, k_2 = 5, N_0 = 2, \tau'_2 = 5, a = 4.
\end{align}
\label{eqn:params-v22-time-progression}
\end{subequations}

\end{appendices}

\addcontentsline{toc}{section}{References}
\bibliographystyle{unsrt}
\bibliography{CC}

\end{document}